\documentclass[10pt,onecolumn,preprintnumbers,amsmath,amssymb,nofootinbib,superscriptaddress,aps,prd,numberedappendix]{openjournal}

\usepackage{newtxtext,newtxmath}
\usepackage{lipsum}
\usepackage[T1]{fontenc}
\usepackage{ae,aecompl}
\usepackage[nodisplayskipstretch]{setspace}

\usepackage{graphicx}	
\usepackage{amsmath}	
\usepackage{amssymb}	

\begin{document}
\vspace{-0.5cm}
\title{The Impact of Quadratic Biases on Cosmic Shear}

\author{T. D. Kitching$^{1,\dagger}$ 
, A. C. Deshpande$^{1}$}
\email{$^{\dagger}$t.kitching@ucl.ac.uk, © 2021. All rights reserved.}
\affiliation{
$^{1}$Mullard Space Science Laboratory, University College London, Holmbury St Mary, Dorking, Surrey RH5 6NT, UK}

\begin{abstract}
In this paper we revisit potential biases in cosmic shear power spectra caused by bias terms that multiply up to quadratic powers of the shear. Expanding the multiplicative bias field as a series of independent spin-$s$ fields we find terms $m_s$ that multiply integer and half-integer powers of the shear. We propagate these biases into shape measurement statistics and the cosmic shear power spectrum. We find that such biases can be measured by performing regression on calibration data. We find that for integer powers of shear the impact of quadratic order terms on the power spectrum is an additional bispectrum dependency; ignoring quadratic terms can lead to biases in cosmological parameters of up to $2(m_2+m_{-2}-m_6)0.4\sigma$ for Stage-IV dark energy experiments, but that the susceptibility to them can be decreased by using methods to remove small-scale sensitivity. We also find, for half-integer powers of the shear that, for a Stage-IV experiment, biases are required to be known to better than approximately $\sigma[m_0+15(m_1+m_3)+0.1(m_{-1}+m_5)]\leq 0.01$. In future, Stage-IV dark energy experiments should seek to measure and minimise such all such bias terms.
\end{abstract}

\maketitle
\vspace{-0.5cm}
\section{Introduction}
\label{S:Intro}
The weak lensing effect on galaxies images is to induce a change in the third eccentricity or third flattening, known as shear. But the inference of shear from data can be biased with respect to the true shear by many effects, for example inaccuracies in the algorithms used to measure galaxy shapes \citep{step1,step2,great08,great10,great3}, the size of the point spread function (PSF)  \citep{2017MNRAS.468.3295H,2019A&A...624A..92K, 2021MNRAS.504.4312G}, detector effects \citep{2014JInst...9C3048A}, or detection effects \citep{cccp, 2021arXiv210810057H}. To linear order the two biases that relate the inferred/measured shear to the true shear are a multiplicative bias (that multiplies the true shear by a value) and an additive bias (that adds a value to the true shear). The propagation of such biases into cosmic shear power spectra is shown in \cite{K19,K20,K21}, that include multiplicative and additive biases and the impact of masked data sets. 

In the majority of weak lensing studies to date \cite[see e.g.][]{great08,great3,kids,DESY3_1} a linear multiplicative bias has been assumed \citep{2005PhRvD..72d3503G, 2006MNRAS.366..101H}. This may appear to be a reasonable approximation, because the amplitude of shear is typically small and so quadratic terms may be even smaller. However, going beyond this assumption \cite{step1} and \cite{great10} both fitted a quadratic term to the available methods at the time, and found values for a coefficient describing this effect of between $0.01-1$, depending on the method in question. In this paper we revisit the impact of such quadratic terms.

As discussed in \cite{K21} the observed shear field, including biases, noise, can be written to linear order, including only integer powers of shear, in multiplicative bias terms like 
\begin{eqnarray}
    \label{gamma1}
    \widetilde\gamma(\mathbf{\Omega})&=&[1+m_0(\mathbf{\Omega})]\gamma(\mathbf{\Omega})+m_4(\mathbf{\Omega})\gamma^*(\mathbf{\Omega})
+c(\mathbf{\Omega}),
\end{eqnarray}
where each of the quantities are dependent on angular coordinates $\mathbf{\Omega}=(\theta,\phi)$, where $\theta$ and $\phi$ are latitude and longitude (or R.A. and dec). The true spin-2 shear (in this nomenclature, spin-$s$ means spin positive $s$) is $\gamma(\mathbf{\Omega})$, and the measured spin-2 shear is $\widetilde\gamma(\mathbf{\Omega})$. $m_0(\mathbf{\Omega})$ and $m_4(\mathbf{\Omega})$ are spin-0 and spin-4 position-dependent multiplicative bias terms respectively. We note here that each term needs to be spin-2, shear is spin-2 so at linear order only a spin-0 multiplicative bias or a spin-4 multiplied by a spin-(-2) field can contribute to linear order. $c(\mathbf{\Omega})$ is a spin-2 position-dependent additive bias. Here we do not include the un-lensed uncorrelated galaxy ellipticity or a survey mask but refer to \cite{K21} for these generalisations; also any intrinsic alignment terms are captured in the $\gamma(\mathbf{\Omega})$ term. $^*$ is a complex conjugate. In general we use $\widetilde{x}$ to mean an observed quantity.

This can be generalised beyond linear order, and beyond integer powers of shear, including new multiplicative bias terms. To do this we create a geometric series of powers of shear that result in integer-spin fields; because shear is spin-2 this means taking either integer or half-integer powers\footnote{Taking any other power would result in fields that were not symmetric under rotations of $2\pi$.}. This results in 
\begin{eqnarray}
    \label{gamma2}
    \widetilde\gamma(\mathbf{\Omega})&=&\gamma(\mathbf{\Omega})+
    \left[\sum_{i=0}^M\sum_{j=0}^N m^{i,j}_{2(1-i/2+j/2)}(\mathbf{\Omega})\gamma^{i/2}(\mathbf{\Omega})\gamma^{*,{j/2}}(\mathbf{\Omega})\right],
\end{eqnarray}
where $i$ and $j$ are integers, and the multiplicative bias sum runs up to order $N$ and $M$ respectively. We do not consider $i<0$ or $j<0$ because $\gamma_i\in [-1,1]$ hence a negative power would lead to divergence as $\gamma_i\rightarrow 0$. The subscript in $m^{i,j}_s(\mathbf{\Omega})$ refers to a series of multiplicative bias terms with spin-$2(1-i/2+j/2)$, where $s=2(1-i/2+j/2)\in \mathbb{Z}$ (i.e. integer); the superscript is a label for the term corresponding to a multiplication of $\gamma^{i/2}(\mathbf{\Omega})\gamma^{*,{j/2}}(\mathbf{\Omega})$, not a power. We note that $c(\mathbf{\Omega})$ corresponds to $m^{0,0}_2(\mathbf{\Omega})$. In creating this geometric expansion terms with power $i/2$ should not be equated with the $(i-2)$ edth derivative of the shear field \citep[see][]{2005PhRvD..72b3516C}, although both do have the same spin-weight. We assume that non-integer powers relate to the positive root of the complex value in question i.e. we assume no additional phases of $\pi(ik)$ where $k \in \mathbb{Z}$ \citep[see][]{markushevich2005theory}, and assume any sign information of each term in the series is captured by the biases $m^{i,j}_s$. In addition to a geometric expansion of shear with spin-weighted bias coefficients (as in equation \ref{gamma2}), one may also want to expand each bias as a function of shear under the assumption that the biases are functions of the shear field $m^{i,j}_s(\gamma,\gamma^*)$ (e.g. as a Taylor expansion), but we leave this for a future analysis.

If we expand equation (\ref{gamma2}) to second order in shear we find that  
\begin{eqnarray}
    \label{gamma2p5}
    \widetilde\gamma(\mathbf{\Omega})&=&\gamma(\mathbf{\Omega})\nonumber\\
    &+&m^{4,0}_{-2}(\mathbf{\Omega})\gamma^2(\mathbf{\Omega})\nonumber\\
    &+&m^{3,0}_{-1}(\mathbf{\Omega})\gamma^{3/2}(\mathbf{\Omega})\nonumber\\
    &+&m^{3,1}_0(\mathbf{\Omega})\gamma^{3/2}(\mathbf{\Omega})\gamma^{*,1/2}(\mathbf{\Omega})\nonumber\\
    &+&m^{2,2}_0(\mathbf{\Omega})\gamma(\mathbf{\Omega})\nonumber\\
    &+&m^{1,0}_1(\mathbf{\Omega})\gamma^{1/2}(\mathbf{\Omega})\nonumber\\
    &+&m^{2,1}_1(\mathbf{\Omega})\gamma(\mathbf{\Omega})\gamma^{*,1/2}(\mathbf{\Omega})\nonumber\\
    &+&m^{2,2}_{2}(\mathbf{\Omega})\gamma(\mathbf{\Omega})\gamma^*(\mathbf{\Omega})\nonumber\\
    &+&m^{1,1}_{2}(\mathbf{\Omega})\gamma^{1/2}(\mathbf{\Omega})\gamma^{*,1/2}(\mathbf{\Omega})\nonumber\\
    &+&m^{0,1}_3(\mathbf{\Omega})\gamma^{*,1/2}(\mathbf{\Omega})\nonumber\\
    &+&m^{1,2}_3(\mathbf{\Omega})\gamma^{1/2}(\mathbf{\Omega})\gamma^{*}(\mathbf{\Omega})\nonumber\\
    &+&m^{0,2}_4(\mathbf{\Omega})\gamma^{*}(\mathbf{\Omega})\nonumber\\
    &+&m^{1,3}_4(\mathbf{\Omega})\gamma^{1/2}(\mathbf{\Omega})\gamma^{*,3/2}(\mathbf{\Omega})\nonumber\\
    &+&m^{0,3}_5(\mathbf{\Omega})\gamma^{*,3/2}(\mathbf{\Omega})\nonumber\\
    &+&m^{0,4}_{6}(\mathbf{\Omega})\gamma^{*,2}(\mathbf{\Omega})\nonumber\\
    &+&c(\mathbf{\Omega}),
\end{eqnarray}
where $m^{i,j}_{2}(\mathbf{\Omega})$ are spin-$2$ multiplicative bias fields and similarly for the other terms. We define quadratic/second order as $\gamma^i\gamma^{*,j}$ where $i+j\leq 2$. For the remainder of this paper we do not include the $\mathbf{\Omega}$ in each equation for clarity, however an angular dependence of each quantity should be assumed unless otherwise stated. Terms above second order $\mathcal{O}(\gamma^3)$ should be small (we note the irony of saying this in a paper that is revisiting $\mathcal{O}(\gamma^2)$ that have been thought to have been unimportant), and we leave investigation of the impact of these (if any exist) to future work. To simplify the analysis for the remainder of this paper we make the assumption that $m^{i,j}_s=m^{k,l}_s$ for any $i,j,k,l$ i.e. that for each spin-weight there is a single multiplicative bias field; this assumption could be dropped in future. In this case we have 
\begin{eqnarray}
    \label{gamma3}
    \widetilde\gamma(\mathbf{\Omega})&=&\gamma(\mathbf{\Omega})\nonumber\\
    &+&m_{-2}(\mathbf{\Omega})\gamma^2(\mathbf{\Omega})\nonumber\\
    &+&m_{-1}(\mathbf{\Omega})\gamma^{3/2}(\mathbf{\Omega})\nonumber\\
    &+&m_0(\mathbf{\Omega})[\gamma(\mathbf{\Omega})+\gamma^{3/2}(\mathbf{\Omega})\gamma^{*,1/2}(\mathbf{\Omega})]\nonumber\\
    &+&m_1(\mathbf{\Omega})[\gamma^{1/2}(\mathbf{\Omega})+\gamma(\mathbf{\Omega})\gamma^{*,1/2}(\mathbf{\Omega})]\nonumber\\
    &+&m_{2}(\mathbf{\Omega})[\gamma(\mathbf{\Omega})\gamma^*(\mathbf{\Omega})+\gamma^{1/2}(\mathbf{\Omega})\gamma^{*,1/2}(\mathbf{\Omega})]\nonumber\\
    &+&m_3(\mathbf{\Omega})[\gamma^{*,1/2}(\mathbf{\Omega})+\gamma^{1/2}(\mathbf{\Omega})\gamma^{*}(\mathbf{\Omega})]\nonumber\\
    &+&m_4(\mathbf{\Omega})[\gamma^{*}(\mathbf{\Omega})+\gamma^{1/2}(\mathbf{\Omega})\gamma^{*,3/2}(\mathbf{\Omega})] \nonumber\\
    &+&m_5(\mathbf{\Omega})\gamma^{*,3/2}(\mathbf{\Omega})\nonumber\\
    &+&m_{6}(\mathbf{\Omega})\gamma^{*,2}(\mathbf{\Omega})\nonumber\\
    &+&c(\mathbf{\Omega}),
\end{eqnarray}
where we drop the superscripts for each field.
We will now consider the integer and half-integer terms in turn, since these have distinct impacts on the power spectrum. This also serves as a pedagogical exploration of the impact of these various different biases. In Section \ref{int} we present how integer quadratic terms can impact shape measurement and the cosmic shear power spectrum; both of which can be understood analytically. In Section \ref{half} we add half-integer quadratic terms and show how then can impact impact shape measurement and the cosmic shear power spectrum. In Section \ref{Discussion} and \ref{Conclusions} we discuss the results and present conclusions.

\section{The Impact of Quadratic-Order Integer Terms on Shape Measurement}
\label{int}
We first consider how biases of integer powers of shear are related to a canonical Cartesian parameterisation of the shear as $\gamma=\gamma_1+{\rm i}\gamma_2$ (where $\gamma_1$ are local distortions parallel to a local axes, and $\gamma_2$ are at $45$ degrees). In this case we find that 
\begin{eqnarray}
    \label{gamma12}
    \Delta\gamma_1=\widetilde\gamma_1-\gamma_1&=&(m^R_0+m^R_4)\gamma_1+(m^R_2+m^R_{-2}+m^R_6)\gamma^2_1+(m^R_2-m^R_{-2}-m^R_6)\gamma^2_2\nonumber\\
    &-&(m^I_0-m^I_4)\gamma_2-2(m^I_{-2}-m^I_6)\gamma_1\gamma_2\nonumber\\
    &+&c_1\nonumber\\
    \Delta\gamma_2=\widetilde\gamma_2-\gamma_2&=&(m^R_0-m^R_4)\gamma_2+2(m^R_{-2}-m^R_6)\gamma_1\gamma_2\nonumber\\
    &+&(m^I_0+m^I_4)\gamma_1+(m^I_2+m^I_{-2}+m^I_6)\gamma^2_1+(m^I_2-m^I_{-2}-m^I_6)\gamma^2_2\nonumber\\
    &+&c_2
\end{eqnarray}   
where we have introduced the notation $\Delta\gamma_i$ to mean the difference between the measured and true shear for component $i$. Each spin-$s$ multiplicative bias field is split into its real and imaginary parts $m_s=m^R_s+{\rm i}m^I_s$ or $m_s=|m_s|{\rm exp}(s\theta_m)$ where $\theta_m$ is the complex angle in the bias' coordinate frame. Note that we consider $m_s$ and $m_{-s}$ to be independent multiplicative bias fields; one could also assume these are related via $m_{-s}=m^*_s$ (and  such a model could be tested against the  calibration data). The inclusion of imaginary parts in the $m_s$ fields allows for a rotation of the observed shear with respect to the true shear\footnote{In \cite{2008MNRAS.389..173K} a different approach was taken where a rotation operator was applied to the shear where $m=|m|{\rm exp}({\rm i}\phi)$, where $\phi$ is an arbitrary rotation angle between the bias coordinate frame and the galaxy image's coordinate frame, so that $\widetilde\gamma=|m||\gamma|{\rm exp}(2{\rm i}(\theta+\phi/2))$, where $\theta={\rm atan}(\gamma_2/\gamma_1)$. Using similar notation, in general the multiplicative terms from equation \ref{gamma2} could be rewritten like 
\begin{eqnarray}
m^{i,j}_{s}\gamma^{i/2}\gamma^{*,{j/2}}{\rm e}^{{\rm i}\phi}=   
|m^{i,j}_{s}||\gamma|^{i/2-j/2}{\rm e}^{2{\rm i}([i/2-j/2]\theta+s\theta/2}){\rm e}^{{\rm i}\phi}=|m^{i,j}_{s}||\gamma|^2{\rm e}^{2{\rm i}\theta}{\rm e}^{{\rm i}\phi},
\end{eqnarray}
here $\phi$ is rotation angle between coordinate frames, $\theta$ is the angle within a coordinate frame, and $s=2(1-i/2+j/2)$. In this case the angle $\phi$ can be incorporated into the multiplicative bias field like 
\begin{eqnarray}
m^{i,j}_{s}=|m^{i,j}_{s}|{\rm e}^{s{\rm i}(\theta+2\phi/s)}. 
\end{eqnarray}
In this case $\theta$ is the angle within the coordinate of the bias field, and $\psi=2\phi/s$ is a rotation angle. It can then be seen that $m^R_s=|m^{i,j}_{s}|\cos(s[\theta+\psi])$ and $m^I_s=|m^{i,j}_{s}|\sin(s[\theta+\psi])$. For  $s\not=0$ a rotation can therefore be incorporated into the multiplicative bias. In this approach we also see that, for $s=0$ one can combine a rotation with the $m_0$ term so that $m^R_0=|m^{i,j}_0|\cos(\psi)$ and $m^I_0=|m^{i,j}_0|\sin(\psi)$ i.e. an `effective' imaginary term that results from combining the scalar and rotational biases, even though $m_0$ is itself a scalar.}. 

One can also make a transformation of variables to $u=(1/2^{1/2})(\gamma_1+\gamma_2)$ and $v=(1/2^{1/2})(\gamma_1-\gamma_2)$, so that equation (\ref{gamma12}) can be rewritten
\begin{eqnarray}
\label{ueq}
    \Delta\gamma_2&=&
    \frac{1}{2^{1/2}}(m^R_0+m^R_4)(u+v)+\frac{1}{2}(m^R_2+m^R_{-2}+m^R_6)(u^2+v^2+2uv)+\frac{1}{2}(m^R_2-m^R_{-2}-m^R_6)(u^2+v^2-2uv)\nonumber\\    
    &-&\frac{1}{2^{1/2}}(m^I_0-m^I_4)(u-v)-(m^I_{-2}-m^I_6)(u^2-v^2)\nonumber\\
    &+&c_1,\nonumber\\
    \Delta\gamma_2&=&\frac{1}{2^{1/2}}(m^R_0-m^R_4)(u-v)+(m^R_{-2}-m^R_6)(u^2-v^2)\nonumber\\
    &+&\frac{1}{2^{1/2}}(m^I_0+m^I_4)(u+v)+\frac{1}{2}(m^I_2+m^I_{-2}+m^I_6)(u^2+v^2+2uv)+\frac{1}{2}(m^I_2-m^I_{-2}-m^I_6)(u^2+v^2-2uv)\nonumber\\
    &+&c_2.
\end{eqnarray}
One may then perform the following regression: $\Delta\gamma_1$ vs. $\gamma_1$; $\Delta\gamma_1$ vs. $\gamma_2$; $\Delta\gamma_2$ vs. $\gamma_1$;  $\Delta\gamma_2$ vs. $\gamma_2$; and also $\Delta\gamma_2$ vs. $u$ (or $v$). By taking derivatives of equation (\ref{gamma12}) we can predict the behaviour of these relationships and how they depend on the multiplicative bias fields. The non-zero derivatives to second order include (we do not list all of these, but they are indicative):
\begin{eqnarray}
    \label{gamma122}
    \frac{\partial\Delta\gamma_1}{\partial\gamma_1}&=&g_{11}=(m^R_0+m^R_4)+2(m^R_2+m^R_{-2}+m^R_6)\gamma_1-2(m^I_{-2}-m^I_6)\gamma_2;\nonumber\\
    \frac{\partial\Delta\gamma_1}{\partial\gamma_2}&=&g_{12}=2(m_2-m_{-2}-m_6)\gamma_2-(m^I_0-m^I_4)-2(m^I_{-2}-m^I_6)\gamma_1;\nonumber\\
    \frac{\partial^2\Delta\gamma_1}{\partial\gamma^2_1}&=&q_{111}=2(m^R_2+m^R_{-2}+m^R_6);\nonumber\\
    \frac{\partial^2\Delta\gamma_1}{\partial\gamma^2_2}&=&q_{122}=2(m^R_2-m^R_{-2}-m^R_6);\nonumber\\
    \frac{\partial^2\Delta\gamma_1}{\partial\gamma_1\partial\gamma_2}&=&q_{112}=-2(m^I_{-2}-m^I_6);
\end{eqnarray}   
for example. We use the notation $g_{ij}$ to mean a first derivative of $\Delta\gamma_i$ with respect to $\gamma_j$ i.e. the gradient $\partial\gamma_i/\partial\gamma_j$, and $q_{ijk}$ to mean a second derivative i.e. the quadratic term $\partial^2\gamma_i/\partial\gamma_j\partial\gamma_k$ (we also will label $q_{iuu}=\partial^2\gamma_i/\partial u^2$, and similar for $v$). By performing regression between all the combinations of $\Delta\gamma_i$ vs. $\gamma_j$ and $u$ (or $v$), one can fit a quadratic function to each of these, and thereby  determine the value of the multiplicative bias terms
\begin{eqnarray}
    \label{gamma123}
    m^R_0&\simeq& \frac{1}{2}[g_{11}+g_{22}];\,\,\,\,\,\,\,m^R_4\simeq\frac{1}{2}[g_{11}-g_{22}];\nonumber\\
    m^R_2&=&\frac{1}{4}[q_{111}+q_{122}];\,\,\,\,\,\,\,m^R_{-2}=\frac{1}{8}[q_{111}-q_{122}+2q_{221}];\nonumber\\
    m^R_{6}&=&\frac{1}{8}[q_{111}-q_{122}-2q_{221}]\nonumber\\
    m^I_0&\simeq& \frac{1}{2}[g_{21}-g_{12}];\,\,\,\,\,\,\,m^I_4\simeq\frac{1}{2}[g_{21}+g_{12}];\nonumber\\
    m^I_2&=&\frac{1}{4}[q_{211}+q_{222}];\,\,\,\,\,\,\,m^I_{-2}=\frac{1}{8}[q_{211}-q_{222}+2q_{112}];\nonumber\\
    m^I_{6}&=&\frac{1}{8}[q_{211}-q_{222}-2q_{112}],
\end{eqnarray}  
where $q_{221}=q_{2uu}-(1/4)(q_{211}+q_{222})$, and $q_{112}=q_{2vv}-(1/4)(q_{111}+q_{122})$. $m^R_0$, $m^R_4$, and $m^I_4$ can only be approximately determined, but can be solved exactly by subtracting off the quadratic terms from equation (\ref{gamma12}) once determined and performing a linear regression. Alternatively a regression can be performed directly on equation (\ref{gamma12}), i.e. not performing regression in the individual sub-spaces, by creating a loss-function function that is 
\begin{eqnarray}
    \label{reges}
    \chi^2=\frac{1}{2}\sum_g\left\{\frac{[\widetilde\Delta\gamma_{1,g}-\Delta\gamma_1(m_s)]^2}{\sigma^2(\widetilde\Delta\gamma_{1,g})}+\frac{[\widetilde\Delta\gamma_{2,g}-\Delta\gamma_2(m_s)]^2}{\sigma^2(\widetilde\Delta\gamma_{1,g})}\right\},
\end{eqnarray}
where $\widetilde\Delta\gamma_{i,g}$ are the measured deviations from calibration data where each simulated galaxy is labelled $g$, and $\Delta\gamma_i(m_s)$ are given in equation (\ref{gamma12}) over the whole ($\gamma_1$, $\gamma_2$) plane. $\sigma^2(\widetilde\Delta\gamma_{i,g})$ are possible uncertainties on the measured deviations. This is simply the $\chi^2$ fit to the data, in future work this could be generalised to more sophisticated and Bayesian approaches. One then can minimise this function with respect to the values of $m_s$. In practice this is the approach that we take in Section \ref{Test on a real method}.

This analysis differs from standard approaches that typically regress $\Delta\gamma_1$ vs. $\gamma_1$ and $\Delta\gamma_2$ vs. $\gamma_2$ and fit a linear relation where $g_{11}=m^{\rm lin}_1$ and $g_{22}=m^{\rm lin}_2$, we will compare our more general analysis with this approach. If one fits a linear relation one should expect that $m^{\rm lin}_1\simeq m_0+m_4$ and $m^{\rm lin}_2\simeq m_0-m_4$. The quadratic parameter $q$ fit to methods in \cite{step1} to $\gamma_1$ is equivalent to $(m^R_0+m^R_{-2}+m^R_6)$. The $q$ in \cite{great10} corresponds to a $\gamma|\gamma|$ term, which can be written as $\gamma(\gamma\gamma^*)^{1/2}=\gamma^{3/2}(\gamma^*)^{1/2}$, which in reference to equation (\ref{gamma2}) corresponds to a $m_0$ term.

Finally, we note that equation (\ref{gamma12}) can be written in a compact form like 
\begin{eqnarray}
\label{Pauli}
    \Delta\gamma_i&=&m^R_0\gamma_i+(-1)^{i-1}m^R_4\gamma_i+\left[\sum^2_{j=1}\sum^2_{k=1}
    [m^R_{2}|2-i|\sigma_{0,jk}+m^R_{-2}\sigma_{5-2i,jk}+m^R_6(-1)^{i-1}\sigma_{5-2i,jk}]\gamma_j\gamma_k\right]\nonumber\\
    &+&(-1)^i\sum^2_{j=1}m^I_0\sigma_{1,ij}\gamma_j+\sum^2_{j=1}m^I_4\sigma_{1,ij}\gamma_j+\left[\sum^2_{j=1}\sum^2_{k=1}
    [m^I_{2}|1-i|\sigma_{0,jk}+m^I_{-2}\sigma_{2i-1,jk}+m^I_6(-1)^{i}\sigma_{2i-1,jk}]\gamma_j\gamma_k\right]\nonumber\\
    &+&c_i,
\end{eqnarray}
where $i={1,2}$. We use Pauli matrices as a convenient indexing scheme, where $\sigma_{x,jk}$ are Pauli matrices with index $jk$, where the sums pick up $\sigma_0=$$\tiny{\left({\begin{matrix}1 & 0 \\ 0 & 1 \end{matrix}}\right)}$ (the identity matrix), $\sigma_1=$$\tiny{\left({\begin{matrix}0 & 1 \\ 1 & 0 \end{matrix}}\right)}$ and $\sigma_3=$$\tiny{\left({\begin{matrix}1 & 0 \\ 0 & -1 \end{matrix}}\right)}$ terms. 

It is important to note that in order to determine if a shape measurement method has quadratic dependence it is not sufficient to fit a linear relation over a very small range of $\gamma_i$ (e.g. $\ll 0.1$); where quadratic terms may not be measurable in a calibration procedure, but nonetheless may be present. In such a case $m_2$, $m_{-2}$, and $m_6$ may still be significantly non-zero (and therefore lead to a misestimation of the power spectrum, see Section \ref{S:Results}), but simply not be well-determined by the calibration approach. In such cases these would be ``hidden'' biases that were unaccounted for. Throughout we use the range of the distribution from \cite{step2} of $|\gamma_i|\leq 0.06$. 

\subsection{Test on Gaussian simulations}
\label{Test on Gaussian simulations}
To investigate the impact of assuming a linear relation between $\Delta\gamma_1$ vs. $\gamma_1$ we generate $10$,$000$ samples from a Gaussian distribution $\gamma_1={\mathcal N}(0,\sigma)$ with mean zero and width $\sigma=0.1$. We then construct $\widetilde\gamma_1=\gamma_1+(m_0+m_4)\gamma_1+(m_2+m_{-2}+m_6)\gamma^2_1+(m_2-m_{-2}-m_6)\gamma^2_2+c_1$ with $m_0=2\times 10^{-3}$, $m_4=0$, $m_2=1\times 10^{-2}$, $m_{-2}+m_6=m_2$ (in order to remove any $\gamma_2$ dependency for this simple test) and $c_1=0$. We then generate $10$,$000$ realisations where $\gamma^R_1=\gamma_1+\mathcal{N}(0,\sigma/\sqrt{N_{\rm sim}})$ this mimic the process of measuring $\Delta\gamma_1$ from $N_{\rm sim}$ simulations, and we generate $10$,$000$ sets of these; we take $N_{\rm sim}=1\times 10^5$. To each realisation we fit functions $\Delta\gamma_1=g_{11}\gamma_1+q_{111}\gamma^2_1+c_1$ and $\Delta\gamma_1=g_{11}\gamma_1+c_1$.

\begin{figure*}
\centering
\includegraphics[width=0.45\columnwidth]{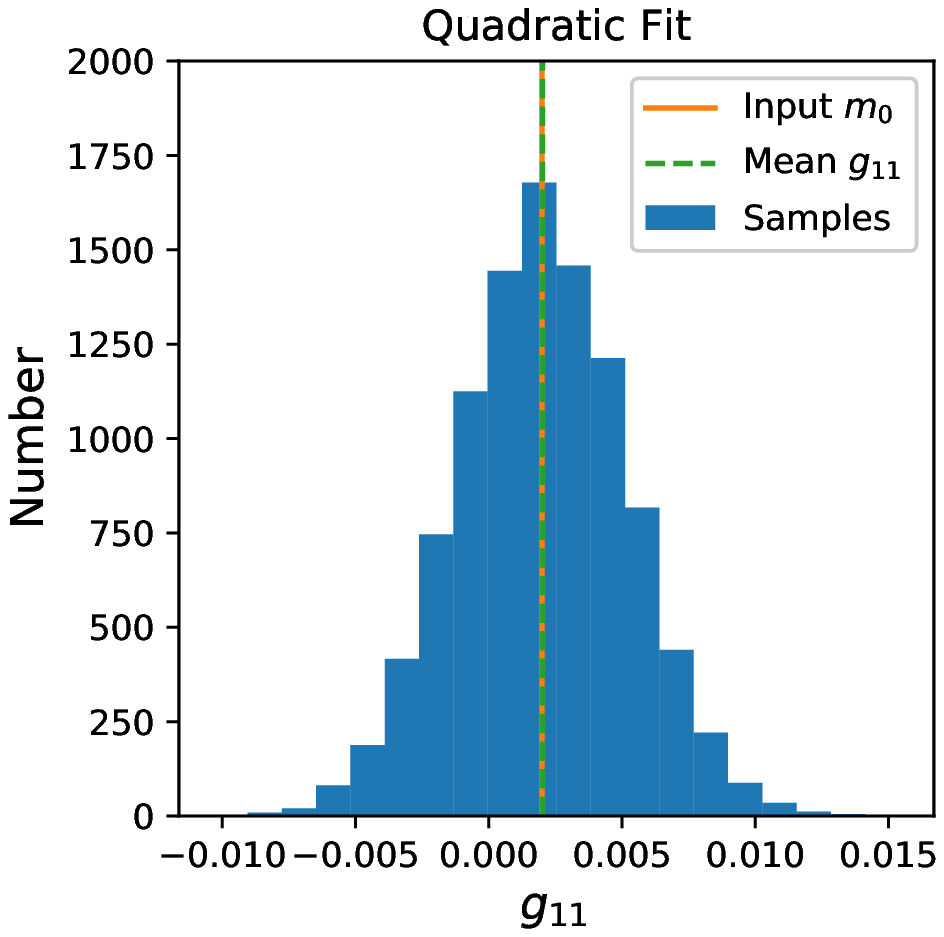}
\includegraphics[width=0.48\columnwidth]{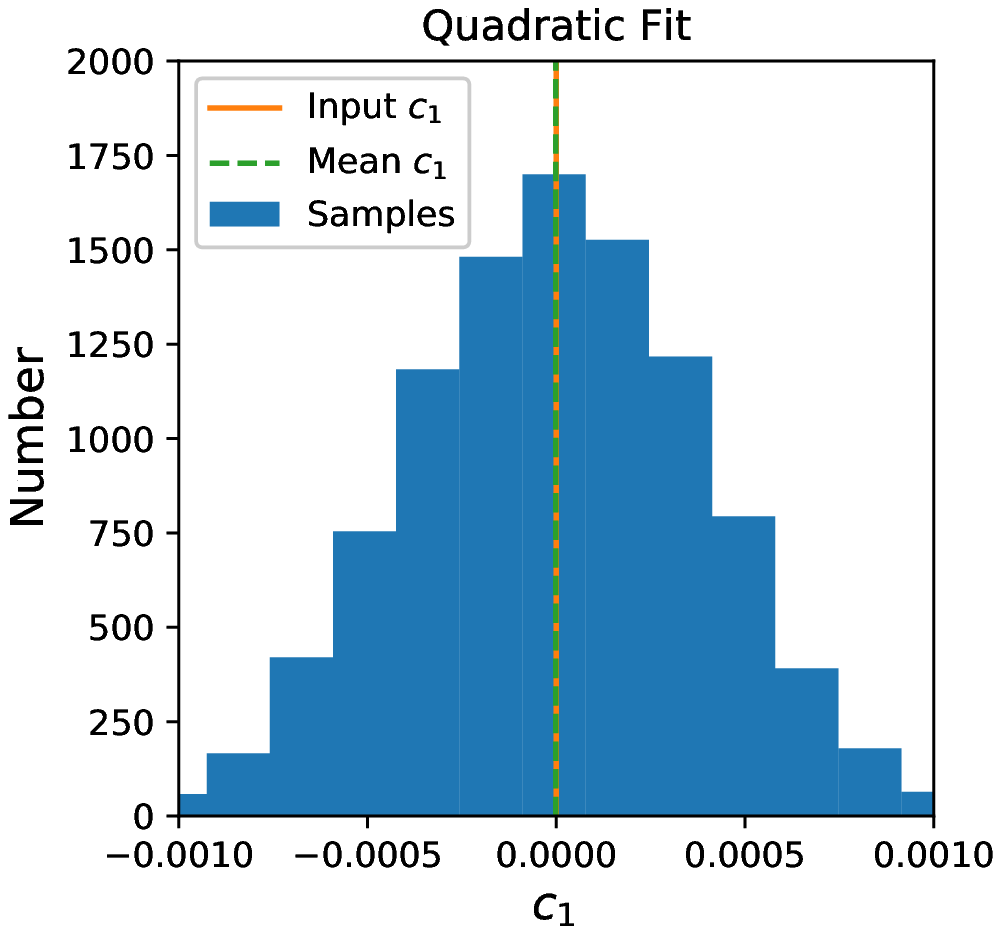}\\
\includegraphics[width=0.45\columnwidth]{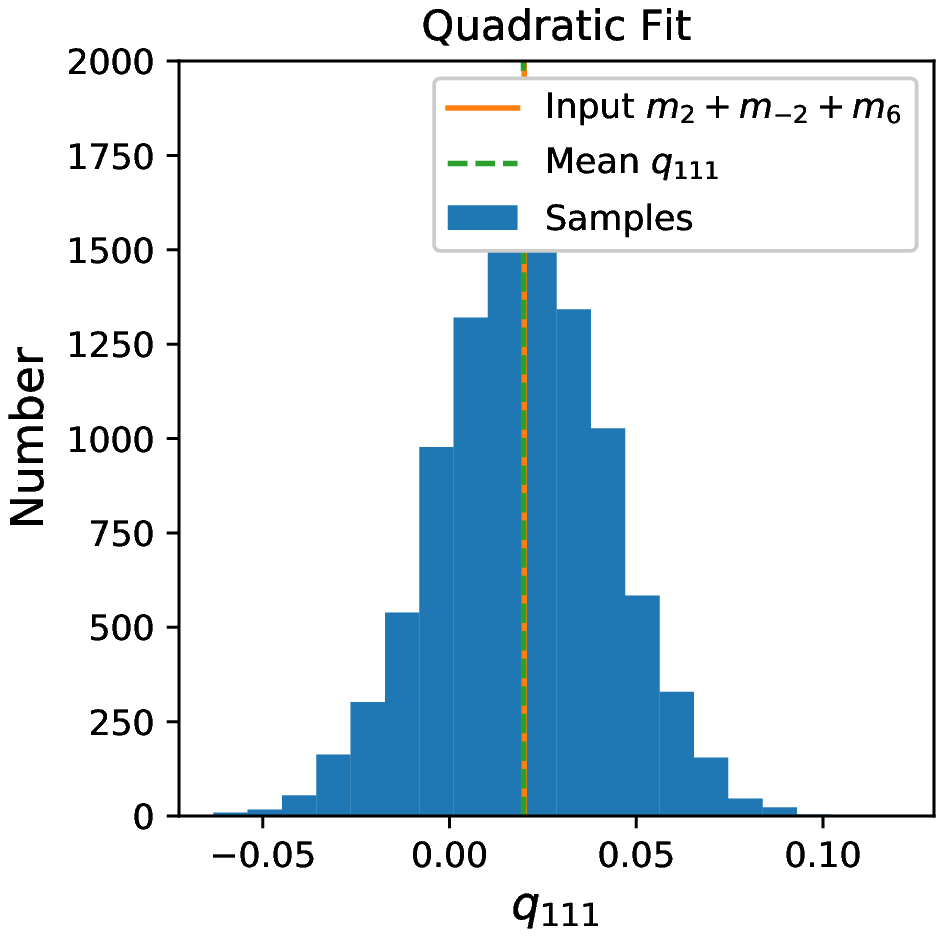}
\includegraphics[width=0.46\columnwidth]{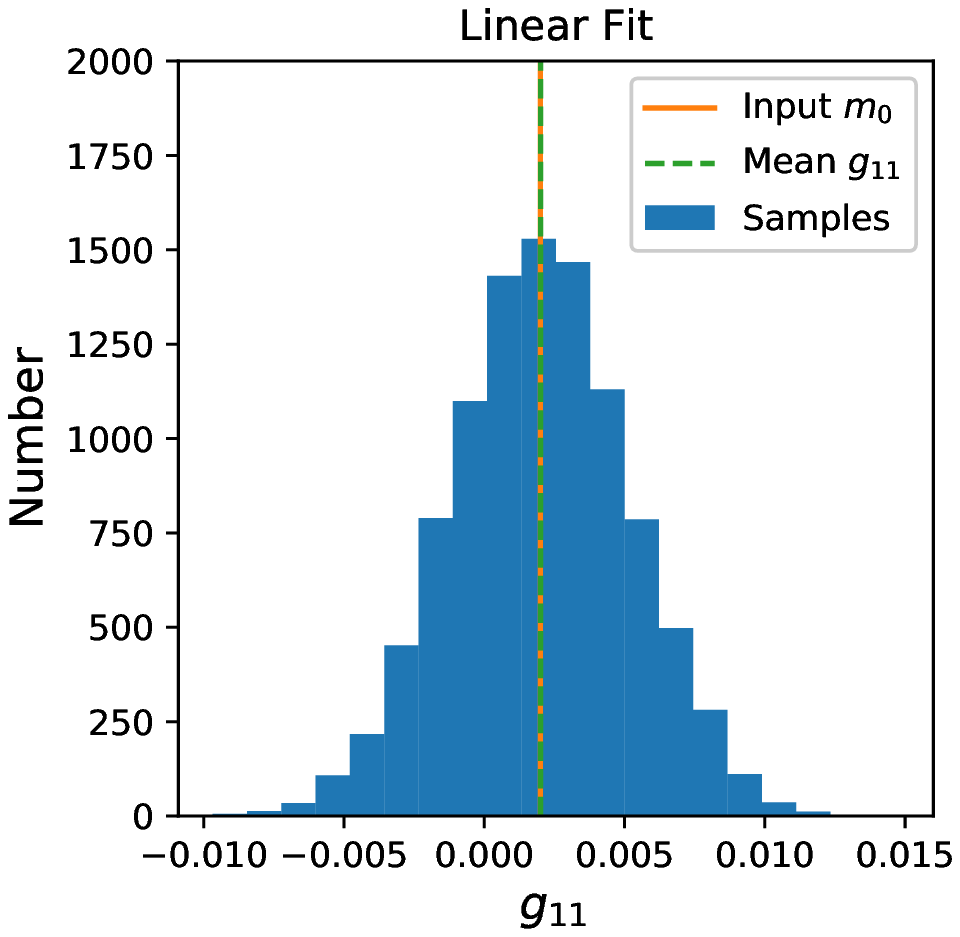}\\
\includegraphics[width=0.45\columnwidth]{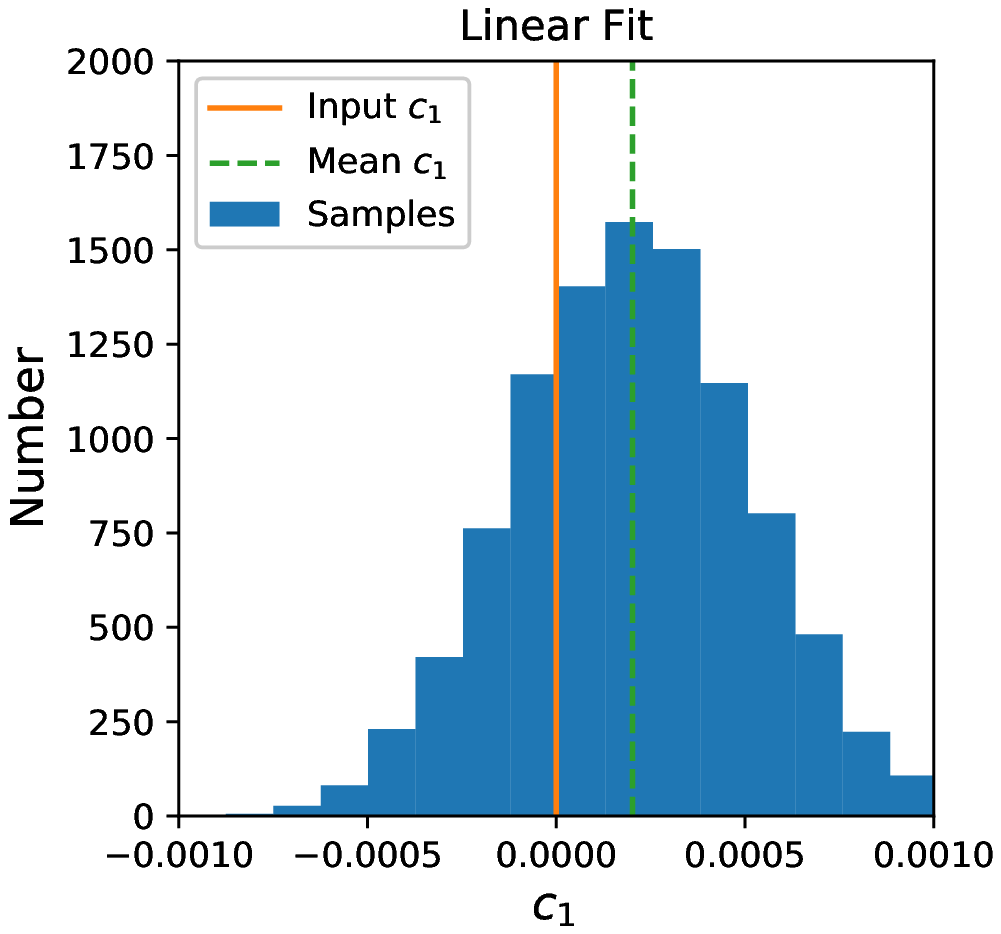}
\includegraphics[width=0.48\columnwidth]{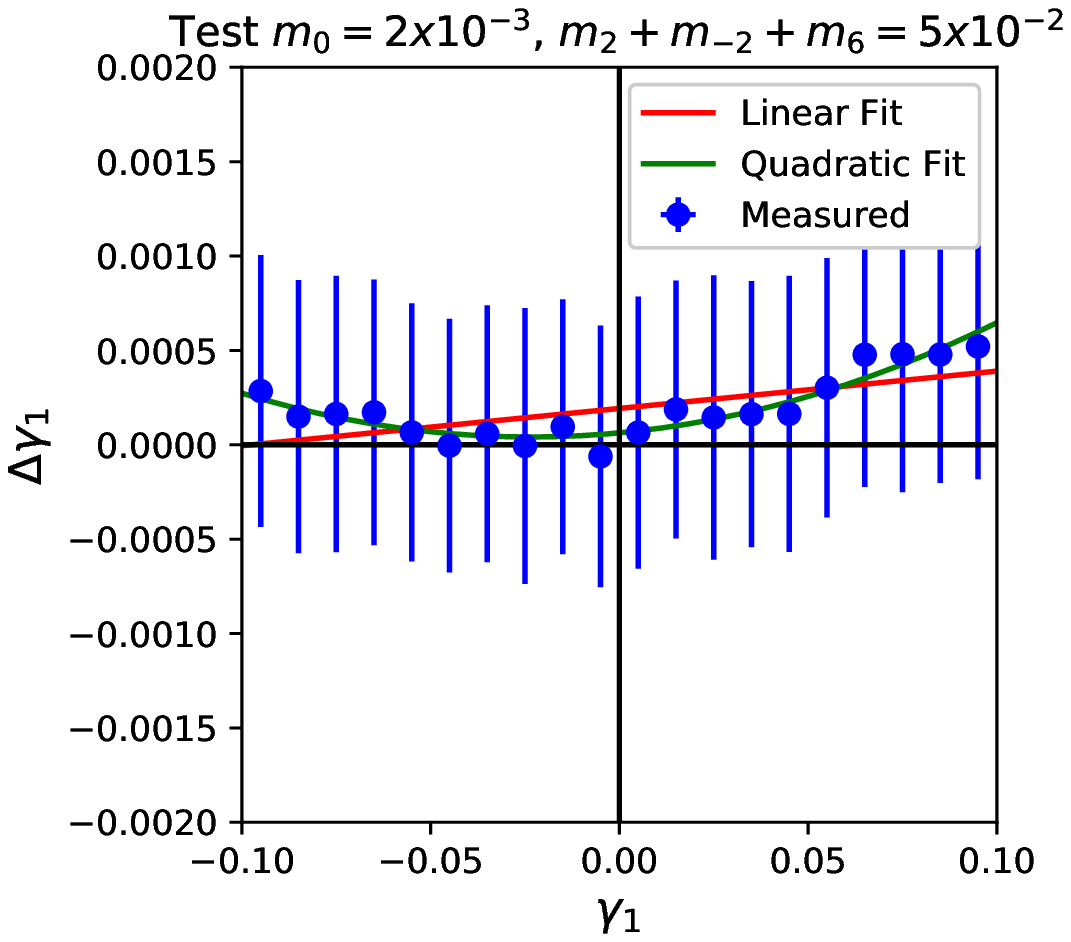}
\caption{A test on randomly generated $\gamma_1$ values ($10$,$000$ realisations of $10$,$000$ galaxies) with $\widetilde\gamma_1=\gamma_1+(m_0+m_4)\gamma_1+(m_2+m_{-2}+m_6)\gamma^2_1+(m_2-m_{-2}-m_6)\gamma^2_2+c_1$ with $\smash{m_0=2\times 10^{-3}}$, $m_4=0$, $\smash{m_2=1\times 10^{-2}}$, $m_{-2}+m_6=m_2$ (in order to remove any $\gamma_2$ dependency for this simple test) and $c_1=0$. Shown are the distribution in the measured values of $g_{11}$, $c_{11}$ and $q_{11}$ assuming a quadratic fit to values $\smash{\Delta\gamma_1=g_{11}\gamma_1+q_{111}\gamma^2_1+c_1}$ (top row), and a linear fit to the values $\Delta\gamma_1=g_{11}\gamma_1+c_1$ (bottom row). The bottom right plot shows an example realisation.} 
\label{Test}
\end{figure*}

\begin{figure}
\centering
\includegraphics[width=0.49\columnwidth]{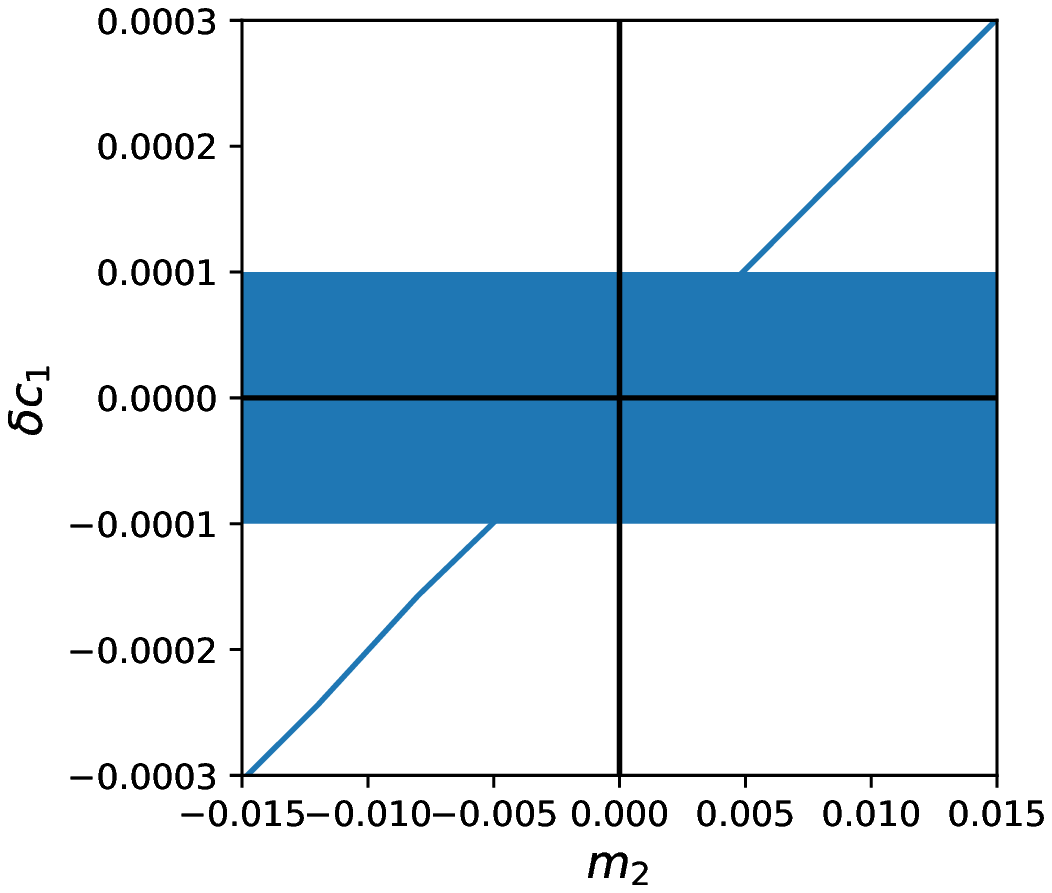}
\includegraphics[width=0.49\columnwidth]{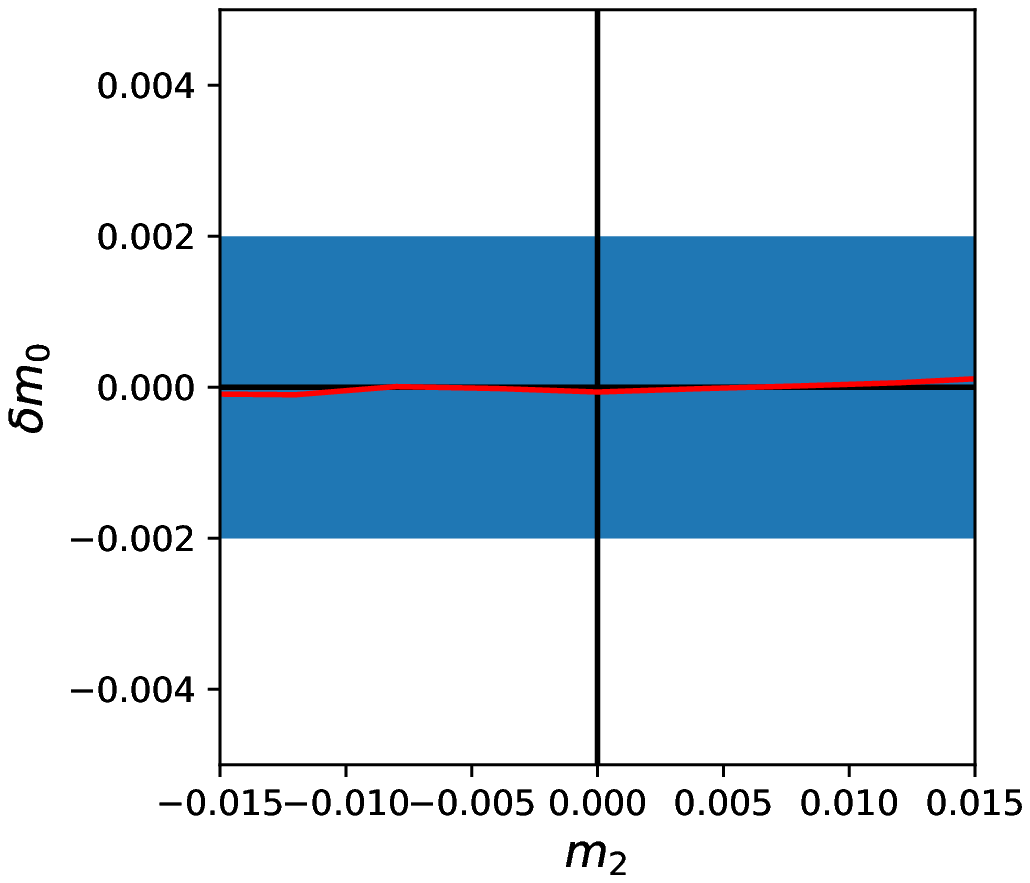}
 \caption{The biased caused in the inferred $c_1$ and $m_0$ caused by a non-zero $m_2$ term, keeping $m_{-2}+m_6=m_2$ and $m_4=0$. The left plot shows the bias in $c_1$ (blue line) and a shaded region $\pm 1\times 10^{-4}$, the right plot shows the bias in $m_0$ (red line) and a shaded region $\pm 2\times 10^{-3}$.} 
\label{qc}
\end{figure}

In Figure \ref{Test} we show the result of these tests. As expected if the correct model is fitted (a quadratic form) then the estimated values for $m_0$, $m_{2}+m_{-2}+m_6$ and $c_1$ are unbiased, however if a linear function is fitted the estimates for $m_0$ and $c_1$ are biased. In particular $c_1$ is biased by $\delta c_1\simeq 2\times 10^{-4}$. In Figure \ref{Test} we also show an example realisation of the $\Delta\gamma_1$ vs. $\gamma_1$ plane in this test setup. In Figure \ref{qc} we vary $m_2$ by up to $1.5\times 10^{-2}$, keeping $m_{-2}+m_6=m_2$, and show how the bias in $c_1$ and $m_0$ vary. We compare these to requirements on $m_0$ and $c_1$ for a Stage-IV \citep{detf} dark energy experiment \citep{cropper}. We find that (under the assumptions of this simple test) if one assumes a linear fit then this introduces a spurious $\delta c_1\simeq 0.02m_2$ whereas the $m_0$ term is relatively unaffected.

\subsection{Test on a real method}
\label{Test on a real method}
To test whether such quadratic terms may be present in a real shape measurement algorithm we use the {\tt GalSim} \citep{galsim} {\tt demo2} which uses a sheared, exponential profile for the galaxy; convolves it by a circular Moffat PSF; and adds Poisson noise to the image. The HSM \citep{HS,M} method is then used to determine the shear values, and we investigate the PSF correction method `KSB' \citep{1995ApJ...449..460K}. We generate $N_{\rm gal}=100$,$000$ realisations where for each we sample the true $\gamma_1$ and $\gamma_2$ from a uniform distribution between $[-0.06, 0.06]$  with $|\gamma|<0.06$. We then minimise the function in equation (\ref{reges}) over the set of $m_s$ values\footnote{This is done using the {\tt scipy.optimize.minimize} function; throughout we use $\sigma(\widetilde\Delta\gamma_{i,g})$ equal to the standard deviation of the distribution of measured $\widetilde\Delta\gamma_{i,g}$ point estimates which is $\sigma(\widetilde\Delta\gamma_{i,g})\simeq 0.01$. The validity of this approach was tested by creating data with known $m_s$ values and checking they were recovered for the noise level matching those in the calibration sample used.}.

\begin{figure*}
\centering
\includegraphics[width=0.49\columnwidth]{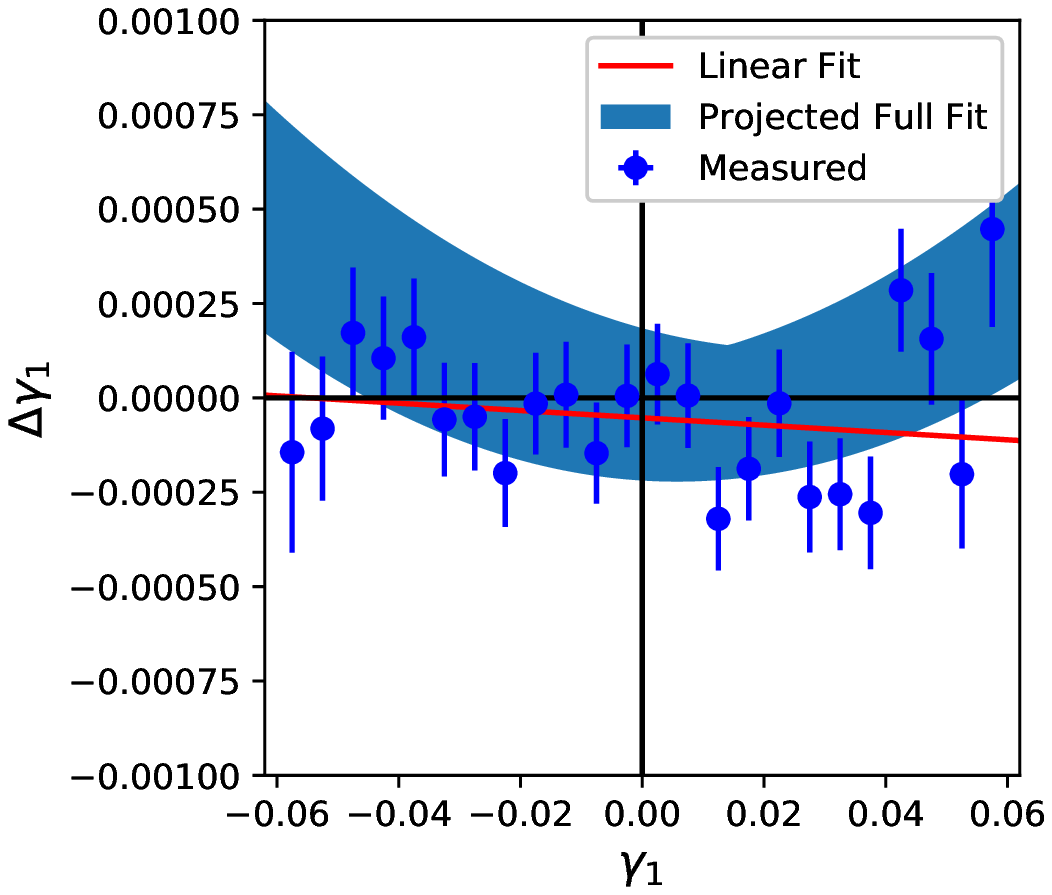}
\includegraphics[width=0.49\columnwidth]{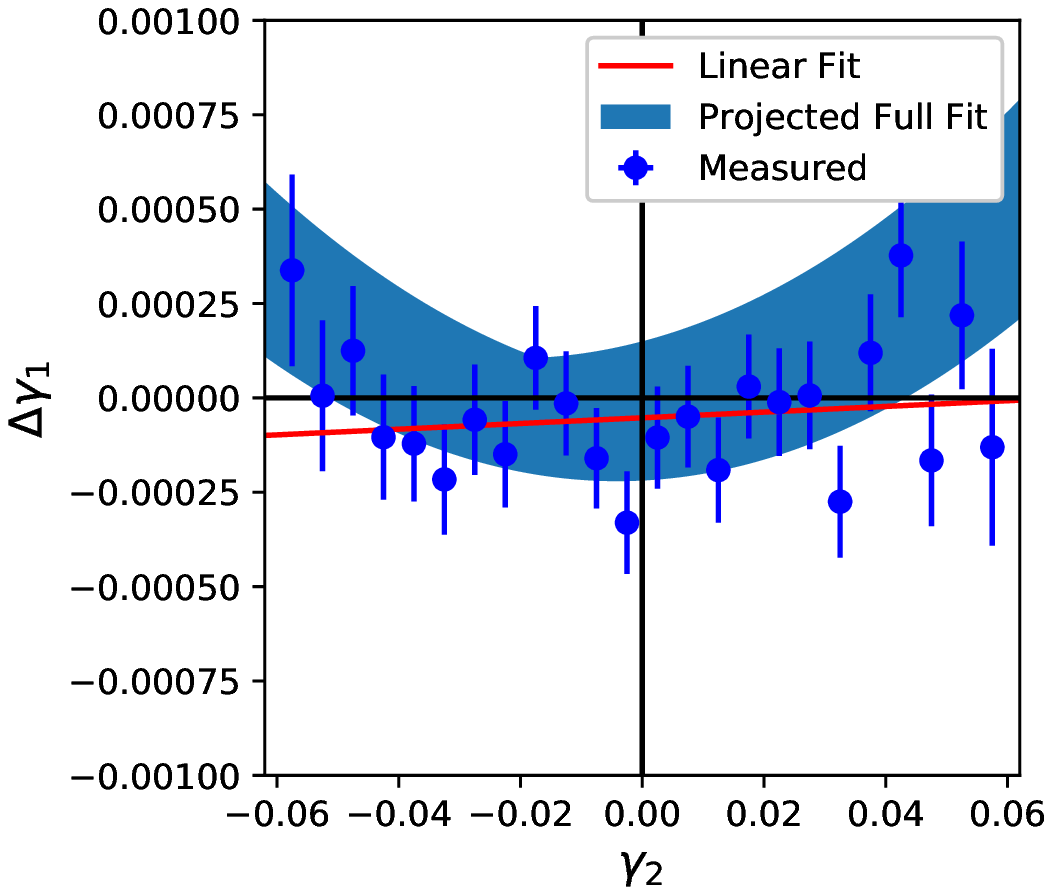}\\
\includegraphics[width=0.49\columnwidth]{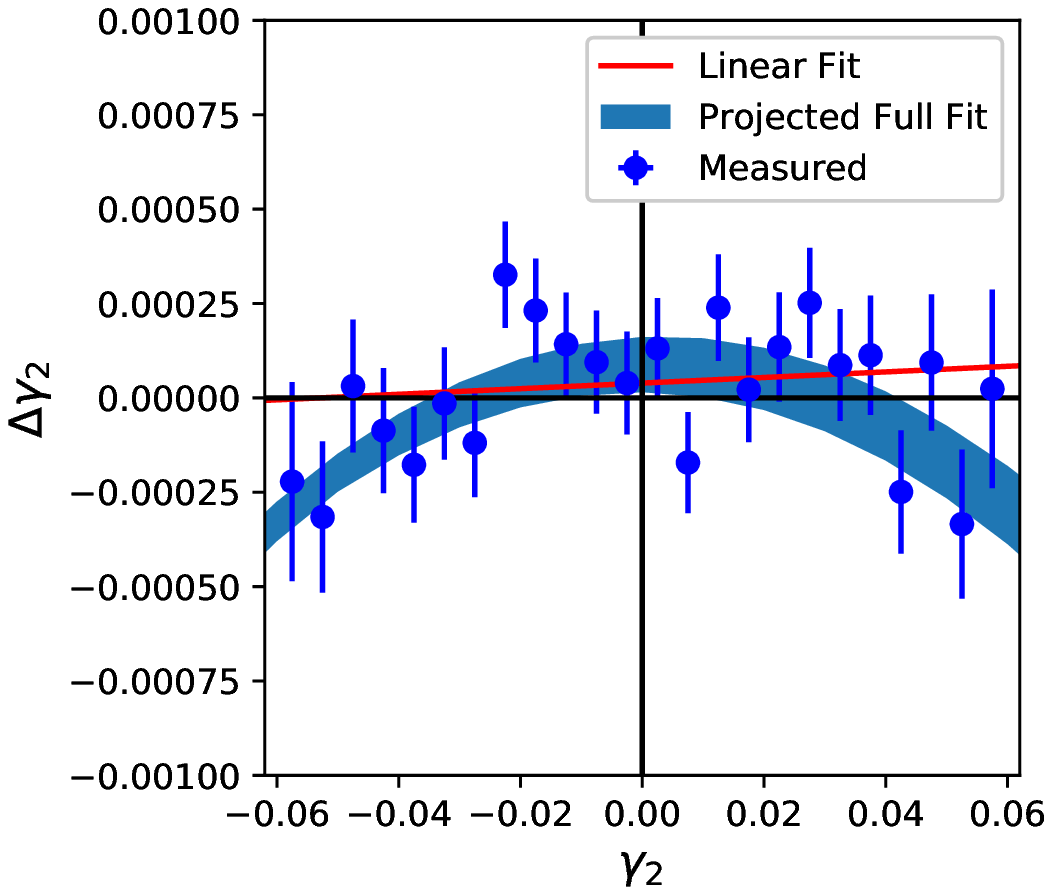}
\includegraphics[width=0.49\columnwidth]{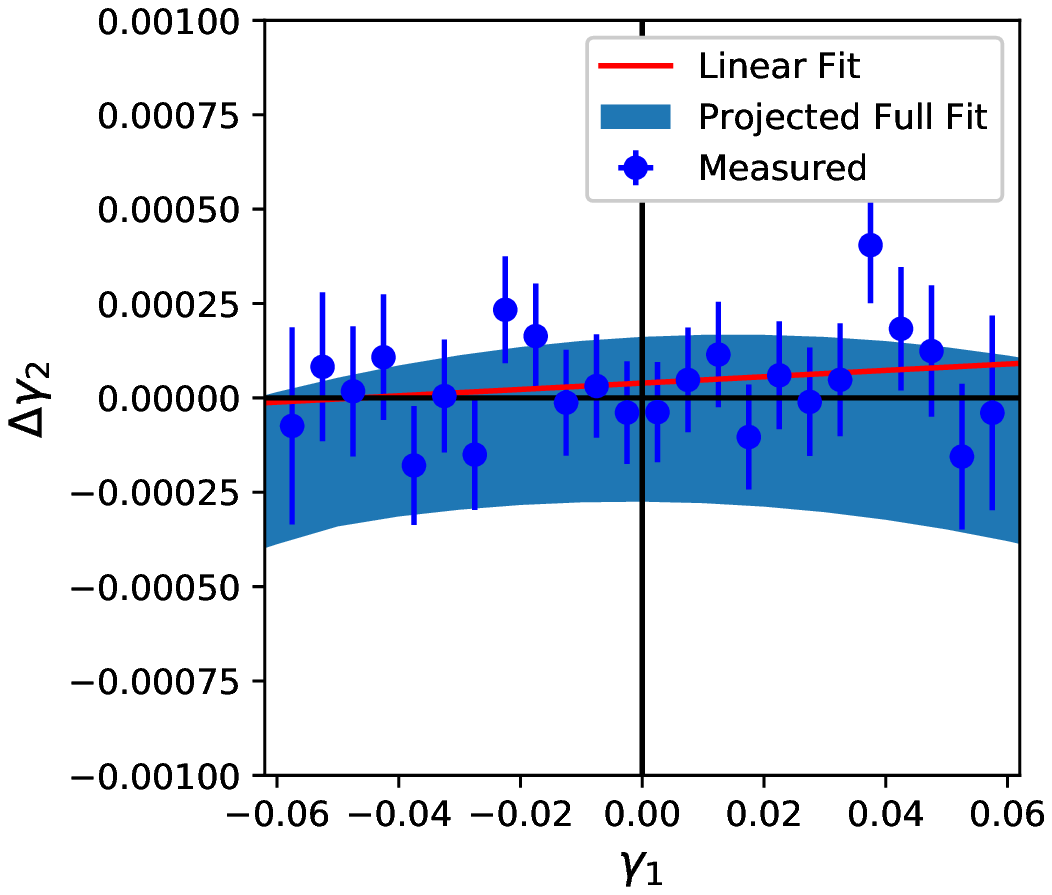}\\
\includegraphics[width=0.49\columnwidth]{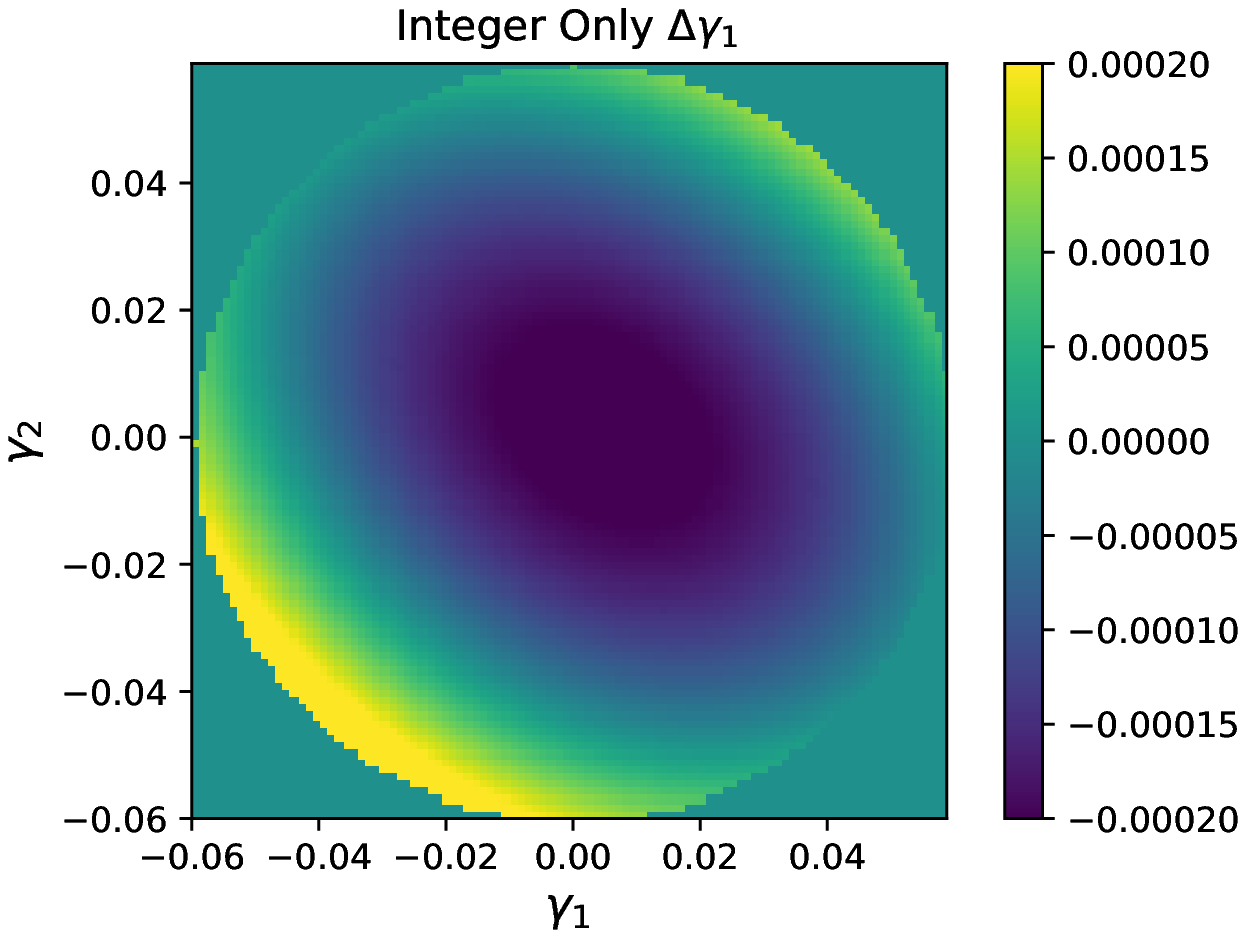}
\includegraphics[width=0.49\columnwidth]{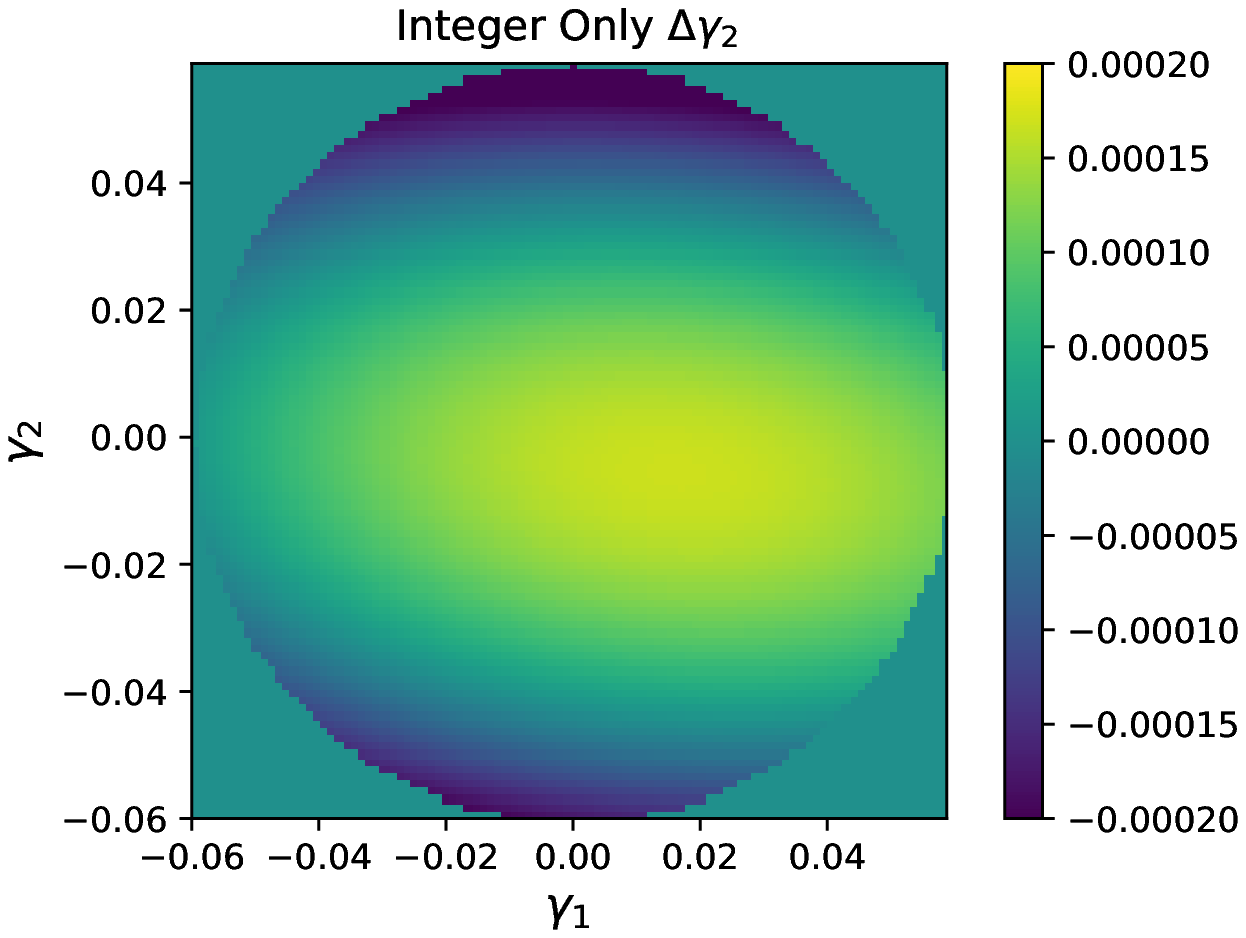}
 \caption{Regression of differences between measured and true shear values in the the $\Delta\gamma_1$ vs. $\gamma_1$; $\Delta\gamma_1$ vs. $\gamma_2$; $\Delta\gamma_2$ vs. $\gamma_2$; and $\Delta\gamma_2$ vs. $\gamma_1$; for the HSM \citep{HS,M} method using the `KSB' PSF correction method; using {\tt GalSim} \citep{galsim} {\tt demo2}. In each plot we show the quadratic fit projected into each plane (blue regions lines), and in all cases a linear fit is shown (red lines). The bottom two plots show the $(\gamma_1,\gamma_2)$ plane with colour showing the amplitude of $\Delta\gamma_1$ (left) and $\Delta\gamma_2$ (right).} 
\label{gs}
\end{figure*}

We show the results in Figure \ref{gs} in the $\Delta\gamma_1$ vs. $\gamma_1$; $\Delta\gamma_1$ vs. $\gamma_2$; $\Delta\gamma_2$ vs. $\gamma_1$;  $\Delta\gamma_2$ vs. $\gamma_2$; and $\Delta\gamma_2$ vs. planes. It is important to appreciate that these planes are projections of the 3D $(\Delta\gamma_i,\gamma_1,\gamma_2)$ fit into 2D planes, therefore in each plot we show the projection of the fitted surface into these planes. We also show the full planes at the bottom of the Figure. We find a reduced $\chi^2_R-1={\rm min}[\chi^2/(N_{\rm gal}-N_{\rm par})]-1=-6\times 10^{-5}$ where $N_{\rm par}=11$.

The individual biases, as well as those found assuming a linear relation are shown in Table 1. The level of spurious additive bias introduced is similar to that found using the simple Gaussian tests. We find that in particular $m_2$ is significantly non-zero. The key feature in these results that indicates non-zero quadratic terms is a non-zero $\Delta\gamma_1$ vs. $\gamma_2$ dependency, which coupled with a near-zero $\Delta\gamma_2$ vs. $\gamma_1$ indicates a significant non-zero $m_2$ term (see equations \ref{gamma12}). These tests are not meant to be a definitive statement on the amplitude of these biases for the HSM (KSB) method, but are meant to demonstrate that in a simple real context such biases are non-negligible.   
\begin{table}[t]
    \centering
    \begin{tabular}{cccc}
    \hline\hline
    Parameter & HSM (KSB)\\
    \hline
    $m^R_0$ & $-0.00011\pm 0.0007$ \\
    $m^I_0$ & $0.00003\pm 0.0007$ \\
    $m^R_4$ & $-0.00078\pm 0.0007$ \\
    $m^R_2$ & $0.031\pm 0.024$ \\
    $m^R_{-2}$ & $-0.0052\pm 0.012$\\
    $m^R_6$ & $-0.00092\pm 0.0012$\\
    $m^I_4$ & $-0.00078\pm 0.0007$ \\
    $m^I_2$ & $-0.022\pm 0.024$ \\
    $m^I_{-2}$ & $0.031\pm 0.012$\\
    $m^I_6$ & $-0.0036\pm 0.0012$\\
    $c_1$ & $-0.00011\pm 0.00004$\\
    $c_2$ &  $0.00013\pm 0.00004$\\
    \hline
    $m^{\rm lin}_1$ & $-0.00096\pm 0.0007$\\
    $m^{\rm lin}_2$ & $0.00074\pm 0.0007$\\
    $c^{\rm lin}_1$ & $0.00005\pm 0.00004$\\
    $c^{\rm lin}_2$ & $0.00004\pm 0.00004$\\
    \hline
    $\delta c_1$ & $-0.00005\pm 0.00004$ \\
    $\delta c_2$ & $0.00009\pm 0.00004$ \\
    \hline\hline
        \label{tab:pars}
    \end{tabular}
    \caption{Measured biases from {\tt GalSim} {\tt demo 2}, using the HSM method with PSF correction `KSB'. The top seven  rows use a quadratic fit over all the $\Delta\gamma_1$ vs. $\gamma_1$; $\Delta\gamma_1$ vs. $\gamma_2$; $\Delta\gamma_2$ vs. $\gamma_1$; $\Delta\gamma_2$ vs. $\gamma_2$ and $\Delta\gamma_2$ vs. $(\gamma_1+\gamma_2)/2^{1/2}$ planes. The middle four rows assume a linear fit only to the $\Delta\gamma_1$ vs. $\gamma_1$; and $\Delta\gamma_2$ vs. $\gamma_2$ planes. The bottom two rows show the different in the additive biases caused by assuming a linear fit relative to the quadratic case.}
\end{table}
\\
\subsection{The Impact of Quadratic-Order Integer Terms on Cosmic Shear Power Spectra}
\label{S:Results}
In this Section we investigate how the presence of quadratic-order integer biases may impact the power spectrum of the shear field. In this derivation we ignore the effect of masks and spatially varying multiplicative bias fields \cite[both of which should be negligible for propagation of such biases into power spectra, see][]{K20,K21}. The two shear components can be combined to extract the E-mode component of the shear field via
\begin{eqnarray}
\widetilde{E}_{\alpha}(\boldsymbol{\ell})=\sum_i T_{i}(\boldsymbol{\ell})\widetilde{\gamma}_{i,\alpha}(\boldsymbol{\ell}).  
\end{eqnarray}
Throughout we use Greek subscripts to label tomographic bins $\alpha$ and $\beta$, and Roman subscripts to denote shear components $i$ and $j$. The trigonometric weighting functions, $T_i(\boldsymbol{\ell})$ are defined as: $T_1(\boldsymbol{\ell}) = \cos(2\phi_\ell)$ and $T_2(\boldsymbol{\ell}) = \sin(2\phi_\ell)$, where $\phi_l$ is the angular component of vector $\boldsymbol{\ell}$ which has magnitude $\ell$. The $E$-mode auto-correlation and cross-correlation spectra, $C_{\ell;\alpha\beta}^{\gamma\gamma}$ can now be defined as:
\begin{equation}
    \label{eq:powerspecdef}
    \widetilde{E}_{\alpha}(\boldsymbol{\ell})\widetilde{E}_{\beta}(\boldsymbol{\ell'}) = (2\pi)^2\delta^2(\boldsymbol{\ell}+\boldsymbol{\ell'})C_{\ell;\alpha\beta}^{\gamma\gamma},
\end{equation}
where $\delta^2$ is the two-dimensional Dirac delta. Assuming the Limber \citep{1953ApJ...117..134L,2017MNRAS.469.2737K,2017JCAP...05..014L}, flat-Universe \citep{2018PhRvD..98b3522T}, reduced shear \citep{ad2}, and flat-sky \citep{10.1046/j.1365-8711.1998.02054.x} approximations, the $C_{\ell;\alpha\beta}^{\gamma\gamma}$ can be related to the three-dimensional matter over-density power spectrum $P_{\delta\delta}(\boldsymbol{k}, \chi)$ by:
\begin{eqnarray}
    \label{eq:Cl}
    C_{\ell;\alpha\beta}^{\gamma\gamma} = \frac{(\ell+2)!}{(\ell-2)!}\frac{1}{\ell^4}\int_0^{\chi_{\rm lim}}{\rm d}\chi\frac{W_{\alpha}(\chi)W_{\beta}(\chi)}{\chi^2} P_{\delta\delta}(\boldsymbol{k}, \chi);
\end{eqnarray}
where
\begin{eqnarray}
\label{Wa}
    W_{\alpha}(\chi) = \frac{3}{2}\Omega_{\rm m}\frac{H_0^2}{c^2}\frac{d_A(\chi)}{a(\chi)}\int_{\chi}^{\chi_{\rm lim}}{\rm d}\chi'n_{\alpha}(\chi')\frac{d_A(\chi'-\chi)}{d_A(\chi')},
\end{eqnarray}
and $\chi$ is comoving distance. \textit{W$_{\alpha}(\chi)$} is the lensing kernel for sources in bin $\alpha$, where $\Omega_{\rm m}$ is the dimensionless present-day matter density of the Universe, $a(\chi)$ is the scale factor of the Universe, $H_0$ is the Hubble constant, $n_{\alpha}(\chi)$ is the probability distribution of galaxies within bin $\alpha$, and $c$ is the speed of light in a vacuum.

To construct the change in the power spectrum caused by the presence of non-zero $m_2$, $m_{-2}$ and $m_6$ multiplicative biases we start with equation (\ref{Pauli}). We will follow the derivation of \cite{2020A&A...636A..95D} (that is for an analogous problem of reduced shear) and throughout we will assume that there is no strong spatial dependence of the multiplicative bias terms \cite[which should be accurate to a good approximation, see][]{K20}; in the following equations $m_s$ should be read as the mean of the real part of the multiplicative bias fields (the imaginary parts will not contribute to the EE power spectrum). We will also only consider terms of cubic order or lower in the lens potential (i.e. that depend on matter over-density power spectrum or bispectrum). As an example of how bispectrum terms arise we consider just $m^R_{-2}$ contribution where from equation (\ref{Pauli}) we have 
\begin{equation}
    \label{eq:gexpan}
    \widetilde\gamma_{i}(\mathbf{\Omega})=\gamma_{i}(\mathbf{\Omega})+\sum^2_{j=1}\sum^2_{k=1}\sigma_{5-2i,jk}m^R_{-2}\gamma_j(\mathbf{\Omega})\gamma_k(\mathbf{\Omega}),
\end{equation}
where $i={1,2}$, and we include the $\mathbf{\Omega}$ explicitly. By computing the power spectrum of this quantity we find the uncorrected power spectrum plus an additional term
\begin{equation}
    \label{eq:ecorr}
     (2\pi)^2\delta^2(\boldsymbol{\ell}+\boldsymbol{\ell'})\delta C^{m_{-2}}_{\ell;\alpha\beta} =  m^R_{-2}\sum_i \sum_j \sum_k \sum_l
     T_i(\boldsymbol{\ell})T_j(\boldsymbol{\ell'})
     \sigma_{5-2i,kl}\langle(\gamma_k\gamma_l)_{\alpha}(\boldsymbol{\ell})
     \gamma_{j,\beta}(\boldsymbol{\ell'})\rangle + T_i(\boldsymbol{\ell'})T_j(\boldsymbol{\ell})\sigma_{5-2i,kl}\langle(\gamma_k\gamma_l)_{\beta}(\boldsymbol{\ell'})\gamma_{j,\alpha}(\boldsymbol{\ell})\rangle,
\end{equation}
where $\delta C^{m_{-2}}_{\ell;\alpha\beta}$ are the resulting corrections to the angular auto and cross-correlation spectra. Applying the Limber approximation, and summing over all terms we obtain an expression:
\begin{equation}
    \label{eq:dCl}
    \delta C^{m^R_{-2}}_{\ell;\alpha\beta} = 2 m_{-2} \left[\frac{(\ell+2)!}{(\ell-2)!}\right]^{3/2}\frac{1}{\ell^6}\int_0^\infty\frac{{\rm d}^2\boldsymbol{\ell'}}{(2\pi)^2}  \cos(2\phi_{\ell'}-2\phi_\ell)B_{\alpha\beta}^{\kappa\kappa\kappa}(\boldsymbol{\ell}, \boldsymbol{\ell'},-\boldsymbol{\ell}-\boldsymbol{\ell'}),
\end{equation}
where the factor $2$ comes from $\sum_{ijk}\sigma_{5-2i,jk}$. 
The factors of $[(\ell+2)!/(\ell-2)!]^{3/2}$ and $1/\ell^6$ arise from foregoing the flat-sky approximation. $B_{\alpha\beta}^{\kappa\kappa\kappa}$, is the two-redshift convergence bispectrum, given by
\begin{align}
    \label{eq:bispecK}
    B_{\alpha\beta}^{\kappa\kappa\kappa}(\boldsymbol{\ell_1}, \boldsymbol{\ell_2}, \boldsymbol{\ell_3}) = \frac{1}{2}\int_0^{\chi_{\rm lim}}\frac{{\rm d}\chi}{\chi^4}W_{\alpha}(\chi)W_{\beta}(\chi) [W_{\alpha}(\chi)+W_{\beta}(\chi)] B_{\delta\delta\delta}(\boldsymbol{k_1},\boldsymbol{k_2},\boldsymbol{k_3},\chi),
\end{align}
where $\boldsymbol{k_x}=(\boldsymbol{\ell_x}+1/2)/\chi$ (for $x=1,2,3$). $B_{\delta\delta\delta}(\boldsymbol{k_1},\boldsymbol{k_2},\boldsymbol{k_3},\chi)$ is the three-dimensional matter over-density bispectrum. 

Going beyond this demonstrative case, by taking the full power spectrum of equation (\ref{gamma3}), and focussing the real parts of the multiplicative bias fields, we find the uncorrected standard terms that only depend on $m^R_0$ and $m^R_4$ plus the following correction
\begin{eqnarray}
    \label{eq:dClall0}
    \delta C_{\ell;\alpha\beta} &=& 2(1+m^R_0)m^R_{-2} \left[\frac{(\ell+2)!}{(\ell-2)!}\right]^{3/2}\frac{1}{\ell^6}\int_0^\infty\frac{{\rm d}^2\boldsymbol{\ell'}}{(2\pi)^2}  \cos(2\phi_{\ell'}-2\phi_\ell)B_{\alpha\beta}^{\kappa\kappa\kappa}(\boldsymbol{\ell}, \boldsymbol{\ell'},-\boldsymbol{\ell}-\boldsymbol{\ell'})\nonumber\\
    &-& 2(1+m^R_0)m^R_{6} \left[\frac{(\ell+2)!}{(\ell-2)!}\right]^{3/2}\frac{1}{\ell^6}\int_0^\infty\frac{{\rm d}^2\boldsymbol{\ell'}}{(2\pi)^2}  \cos(2\phi_{\ell'}-2\phi_\ell)B_{\alpha\beta}^{\kappa\kappa\kappa}(-\boldsymbol{\ell}, -\boldsymbol{\ell'},-\boldsymbol{\ell}-\boldsymbol{\ell'})\nonumber\\
     &+& 2(1+m^R_0)m^R_{2} \left[\frac{(\ell+2)!}{(\ell-2)!}\right]^{3/2}\frac{1}{\ell^6}\int_0^\infty\frac{{\rm d}^2\boldsymbol{\ell'}}{(2\pi)^2}  \cos(2\phi_{\ell'}-2\phi_\ell)B_{\alpha\beta}^{\kappa\kappa\kappa}(\boldsymbol{\ell}, -\boldsymbol{\ell'},-\boldsymbol{\ell}-\boldsymbol{\ell'})\nonumber\\
       &-& 2m^R_4m^R_{-2} \left[\frac{(\ell+2)!}{(\ell-2)!}\right]^{3/2}\frac{1}{\ell^6}\int_0^\infty\frac{{\rm d}^2\boldsymbol{\ell'}}{(2\pi)^2}  \cos(2\phi_{\ell'}-2\phi_\ell)B_{\alpha\beta}^{\kappa\kappa\kappa}(\boldsymbol{\ell}, \boldsymbol{\ell'},\boldsymbol{\ell}+\boldsymbol{\ell'})\nonumber\\
        &+& 2m^R_4m^R_{6} \left[\frac{(\ell+2)!}{(\ell-2)!}\right]^{3/2}\frac{1}{\ell^6}\int_0^\infty\frac{{\rm d}^2\boldsymbol{\ell'}}{(2\pi)^2}  \cos(2\phi_{\ell'}-2\phi_\ell)B_{\alpha\beta}^{\kappa\kappa\kappa}(-\boldsymbol{\ell}, -\boldsymbol{\ell'},\boldsymbol{\ell}+\boldsymbol{\ell'})\nonumber\\
         &-& 2m^R_4m^R_{2} \left[\frac{(\ell+2)!}{(\ell-2)!}\right]^{3/2}\frac{1}{\ell^6}\int_0^\infty\frac{{\rm d}^2\boldsymbol{\ell'}}{(2\pi)^2}  \cos(2\phi_{\ell'}-2\phi_\ell)B_{\alpha\beta}^{\kappa\kappa\kappa}(\boldsymbol{\ell}, -\boldsymbol{\ell'},\boldsymbol{\ell}+\boldsymbol{\ell'}).
\end{eqnarray}
The varying prefactors come from the sums over the prefectors associated in equation (\ref{Pauli}), and the differing signs in convergence bispectrum occur due to the various combinations of complex conjugation that result from the two-point correlation of equation (\ref{gamma3}), making use of the fact that $\kappa^*(\boldsymbol{\ell})=\kappa(-\boldsymbol{\ell})$ since $\kappa$ is a real field \citep[see e.g.][]{2005A&A...431....9S}. Note that the imaginary terms do not contribute to the EE power spectrum.

To first order in multiplicative bias terms, and assuming the $\ell$-mode dependent prefactor is unity, this reduces to  
\begin{eqnarray}
    \label{eq:dClall}
    \delta C_{\ell;\alpha\beta} &\simeq& 2(m^R_{-2})\int_0^\infty\frac{{\rm d}^2\boldsymbol{\ell'}}{(2\pi)^2}  \cos(2\phi_{\ell'}-2\phi_\ell)B_{\alpha\beta}^{\kappa\kappa\kappa}(\boldsymbol{\ell}, \boldsymbol{\ell'},-\boldsymbol{\ell}-\boldsymbol{\ell'})\nonumber\\
    &-& 2(m^R_{6}) \int_0^\infty\frac{{\rm d}^2\boldsymbol{\ell'}}{(2\pi)^2}  \cos(2\phi_{\ell'}-2\phi_\ell)B_{\alpha\beta}^{\kappa\kappa\kappa}(-\boldsymbol{\ell}, -\boldsymbol{\ell'},-\boldsymbol{\ell}-\boldsymbol{\ell'})\nonumber\\
     &+& 2(m^R_{2})\int_0^\infty\frac{{\rm d}^2\boldsymbol{\ell'}}{(2\pi)^2}  \cos(2\phi_{\ell'}-2\phi_\ell)B_{\alpha\beta}^{\kappa\kappa\kappa}(\boldsymbol{\ell}, -\boldsymbol{\ell'},-\boldsymbol{\ell}-\boldsymbol{\ell'}).
\end{eqnarray}
It can also be shown the integral over the second and third terms in equation (\ref{eq:dClall}) is in fact equal to the first term; because the only angular dependence comes from the cosine in the outermost integral, and since $\cos(2(\phi_{\ell} - [\phi_{\ell'}+\pi]))=\cos(2([\phi_{\ell}+\pi] - [\phi_{\ell'}+\pi]))=\cos(2(\phi_{\ell} -\phi_{\ell'}))$ \citep[there is no angular dependence in the matter bispectrum because there is only one valid closed triangle for any given combination of $\ell$ sides, so the angles between the sides are always the same; see e.g.][for details]{2020ApJ...895..113T}. Therefore the final expression, to linear order in $m_s$ is 
\begin{eqnarray}
    \label{eq:dClall2}
    \delta C_{\ell;\alpha\beta} \simeq 2[m^R_2+m^R_{-2}-m^R_6]\int_0^\infty\frac{{\rm d}^2\boldsymbol{\ell'}}{(2\pi)^2}  \cos(2\phi_{\ell'}-2\phi_\ell)B_{\alpha\beta}^{\kappa\kappa\kappa}(\boldsymbol{\ell}, \boldsymbol{\ell'},-\boldsymbol{\ell}-\boldsymbol{\ell'}).
\end{eqnarray}
This is equal to the reduced shear approximation \citep{2020A&A...636A..95D}, except with a prefactor $2(m^R_2+m^R_{-2}-m^R_6)$. Therefore the expected bias on cosmological parameters caused by ignoring quadratic terms in multiplicative biases should be $\delta\theta_i=2(m^R_2+m^R_{-2}-m^R_6)\delta\theta^{\rm RS}_i$ where $\delta\theta_i$ is a bias in cosmological parameter $\theta_i$, and $\delta\theta^{\rm RS}_i$ are the biases caused by the reduced shear approximation, that are given in \citep{2020A&A...636A..95D} for a \emph{Euclid}-like Stage-IV dark energy experiment. 

The biases on cosmological parameters derived in \citep{2020A&A...636A..95D}, relative to their uncertanties (denoted $\sigma$) are $\delta\Omega_{\rm m}=-0.10\sigma$, $\delta\Omega_{\rm b}=0.023\sigma$, $\delta h=0.072\sigma$, $\delta n_{\rm s}=-0.10\sigma$, $\delta\sigma_8=0.055\sigma$; all of which are relatively small. However, $\delta\Omega_{\rm DE}=0.31\sigma$, $\delta w_0=-0.32\sigma$, and $\delta w_a=0.40\sigma$, all of which are significant biases. We refer to \cite{2020A&A...636A..95D} for the full details of these calculations. 

A reasonable requirement is $\delta\theta_i/\sigma\simeq 0.25$, to ensure 90\% overlap between biased and unbiased confidence regions \citep[see e.g.][]{massey}. Therefore if one wishes the contribution from quadratic terms to be sub-dominant to other sources of potential bias, and only account for $Q\%$ of the total budget then a requirement can be set on the sum of the quadratic multiplicative biases of $|m_2+m_{-2}-m_6| < Q (0.25/0.4)/2=0.3125Q$ (taking the largest of the potential biases, $\delta w_a=0.40\sigma$). If one chooses a reasonable $Q=0.1$, this leads to a requirement that $|m_2+m_{-2}-m_6|<0.031$. If one wishes $Q$ be much smaller this would lead to a proportionately smaller requirement.

\subsection{Discussion of Integer Terms} 
\label{Discussion}
By combining the result from the previous Sections we therefore see that there are two effects of ignoring quadratic biases of integer power in shear. Firstly that spurious additive biases may be present where $\delta c_1\simeq 0.02m_2$ for example. Because the requirement for Stage-IV surveys is on the error on the additive bias where $\sigma(\delta c_i)\leq 10^{-4}$ \citep{massey, cropper, K21} this implies that $\sigma(m_s)\simeq 5\times 10^{-3}$. The exact impact of such a spurious additive term will depend on the spatial variation of the spurious additive bias field \citep[as shown in][]{K21} with $\delta C_{\ell;ij}=2C^{\gamma\delta c}_{\ell;ij}+C^{\delta c\delta c}_{\ell;ij}$, where the autocorrelation of the additive bias and the cross-correlation with the  shear field contribute to the change in the power spectrum. Secondly, ignoring quadratic dependence can bias cosmological parameters by up to $2(m^R_2+m^R_{-2}-m^R_6)0.4\sigma$. A reasonable requirement on the amplitude of these terms is $|m^R_2+m^R_{-2}-m^R_6|\leq 0.031$. However, if $m_2$, $m_{-2}$ and $m_6$ are significantly non-zero, and cannot be reduced, but are known accurately then the bispectrum could be included in any inference and the values of $m^R_2$ $m^R_{-2}$ and $m^R_6$ marginalised over with some associated priors. 

To mitigate the impact of these biases it should be noted that methods to remove small-scale sensitivity from the cosmic shear power spectrum, e.g. k-cut (or x-cut) cosmic shear \citep{2018PhRvD..98h3514T, 2020arXiv200700675T}, can be used to remove susceptibility to reduced shear corrections that have the same scale-dependent form as quadratic bias corrections. Using k-cut cosmic shear \cite{2020PhRvD.102h3535D} find that reduced shear corrections can be removed at the expense of only a 10\% increase in $1\sigma$ error bars on cosmological parameters. If k-cut cosmic shear is applied this increase in error would be the total for both effects (i.e. 10\% total, not 20\%). This worst-case (removal of problematic scales) should not adversely impact the objective for Stage-IV experiments to constrain dark energy parameters to have a Figure-of-Merit of more than $400$ \citep{2013LRR....16....6A,2020A&A...642A.191E}.  

Combining both effects the overall change in the power spectrum caused by ignoring quadratic terms will be
\begin{eqnarray}
    \label{eq:dClall3}
    \delta C_{\alpha\beta} \simeq 2C^{\gamma\delta c}_{\ell;\alpha\beta}+C^{\delta c\delta c}_{\ell;\alpha\beta}-2[m^R_2+m^R_{-2}-m^R_6]\int_0^\infty\frac{{\rm d}^2\boldsymbol{\ell'}}{(2\pi)^2}  \cos(2\phi_{\ell'}-2\phi_\ell)B_{\alpha\beta}^{\kappa\kappa\kappa}(\boldsymbol{\ell}, \boldsymbol{\ell'},-\boldsymbol{\ell}-\boldsymbol{\ell'}),
\end{eqnarray}
where any additional spurious additive bias causes the first two terms (the first of which may be negative), and the third is caused by misestimation of the power assuming $m_2=m_{-2}=m_6=0$.

\section{Including the Impact of Quadratic-Order Half-Integer Terms on Shape Measurement}
\label{half}
We next turn to the half-integer terms in equation (\ref{gamma3}). In general these are more complex to understand because they do not propagate into shape measurement calibration statistics or the power spectrum in a straightforward manner. If we consider the $m_1\gamma^{1/2}$ term and express this in $\gamma_1$ and $\gamma_2$ we find for example that 
\begin{eqnarray}
    m_1\gamma^{1/2}=m_1\left[\frac{(\gamma^2_1+\gamma^2_2)^{1/4}}{[(\gamma_1+(\gamma^2_1+\gamma^2_2)^{1/2})^2+\gamma^2_2]^{1/2}}\right](\gamma_1+(\gamma^2_1+\gamma^2_2)^{1/2}+{\rm i}\gamma_2).
\end{eqnarray}
Hence, unlike the integer terms the analytic propagation of such biases $\Delta\gamma_1$ and $\Delta\gamma_2$ is not trivial. However, the prefactor $(\gamma^2_1+\gamma^2_2)^{1/4}/[(\gamma_1+(\gamma^2_1+\gamma^2_2)^{1/2})^2+\gamma^2_2]^{1/2}$ is approximately constant over the range $|\gamma|<0.06$, with median value of $f\simeq 5.4$. Therefore, a rule-of-thumb is that $m_1\gamma^{1/2}\approx fm_1\gamma_1+{\rm i}fm_1\gamma_2 + f|\gamma|m_1$, hence $m_1$ introduces both a $m_0$-like term and an additive bias.

We can write the change in $\gamma_1$ and $\gamma_2$ including such terms as
\begin{eqnarray}
\label{eeqfullfit}
    \Delta\gamma_1=\widetilde\gamma_1-\gamma_1&=&(m^R_0+m^R_4)\gamma_1+(m^R_2+m^R_{-2}+m^R_6)\gamma^2_1+(m^R_2-m^R_{-2}-m^R_6)\gamma^2_2\nonumber\\
    &-&(m^I_0-m^I_4)\gamma_2-2(m^I_{-2}-m^I_6)\gamma_1\gamma_2\nonumber\\
    &+&c_1\nonumber\\
    &+&(m^R_1+m^R_3)\mathbb{R}(\gamma^{1/2})+m^R_2\mathbb{R}(\gamma^{1/2}\gamma^{*,1/2})+m^R_0\mathbb{R}(\gamma^{3/2}\gamma^{*,1/2})+m^R_4\mathbb{R}(\gamma^{1/2}\gamma^{*,3/2})\nonumber\\
    &+&m^R_1\mathbb{R}(\gamma\gamma^{*,1/2})+m^R_3\mathbb{R}(\gamma^{1/2}\gamma^{*})+m^R_{-1}\mathbb{R}(\gamma^{3/2})+m^R_5\mathbb{R}(\gamma^{*,3/2})\nonumber\\
    &-&(m^I_1-m^I_3)\mathbb{I}(\gamma^{1/2})-m^I_2\mathbb{I}(\gamma^{1/2}\gamma^{*,1/2})-m^I_4\mathbb{I}(\gamma^{1/2}\gamma^{*,3/2})\nonumber\\
    &-&m^I_1\mathbb{I}(\gamma\gamma^{*,1/2})-m^I_3\mathbb{I}(\gamma^{1/2}\gamma^{*})+-m^I_{-1}\mathbb{I}(\gamma^{3/2})-m^I_5\mathbb{I}(\gamma^{*,3/2})\nonumber\\
    \Delta\gamma_2=\widetilde\gamma_2-\gamma_2&=&(m^R_0-m^R_4)\gamma_2+2(m^R_{-2}-m^R_6)\gamma_1\gamma_2\nonumber\\
    &+&(m^I_0+m^I_4)\gamma_1+(m^I_2+m^I_{-2}+m^I_6)\gamma^2_1+(m^I_2-m^I_{-2}-m^I_6)\gamma^2_2\nonumber\\
    &+&c_2\nonumber\\
    &+&(m^R_1-m^R_3)\mathbb{I}(\gamma^{1/2})+m^R_2\mathbb{I}(\gamma^{1/2}\gamma^{*,1/2})+m^R_0\mathbb{I}(\gamma^{3/2}\gamma^{*,1/2})+m^R_4\mathbb{I}(\gamma^{1/2}\gamma^{*,3/2})\nonumber\\
    &+&m^R_1\mathbb{I}(\gamma\gamma^{*,1/2})+m^R_3\mathbb{I}(\gamma^{1/2}\gamma^{*})+m^R_{-1}\mathbb{I}(\gamma^{3/2})+m^R_5\mathbb{I}(\gamma^{*,3/2})\nonumber\\
    &+&(m^I_1+m^I_3)\mathbb{R}(\gamma^{1/2})+m^I_2\mathbb{R}(\gamma^{1/2}\gamma^{*,1/2})+m^I_4\mathbb{R}(\gamma^{1/2}\gamma^{*,3/2})\nonumber\\
    &+&m^I_1\mathbb{R}(\gamma\gamma^{*,1/2})+m^I_3\mathbb{R}(\gamma^{1/2}\gamma^{*})+m^I_{-1}\mathbb{R}(\gamma^{3/2})+m^I_5\mathbb{R}(\gamma^{*,3/2})
\end{eqnarray}
where $\mathbb{R}$ and $\mathbb{I}$ refer to the real and imaginary parts of complex numbers respectively. 

In total, to second order in shear, this results in $20$ free parameters consisting of the real and imaginary parts of the sequence of multiplicative biases $m_s$ with $s=-2$--$5$, plus the real and imaginary parts of $c$.

\begin{figure*}
\centering
\includegraphics[width=0.49\columnwidth]{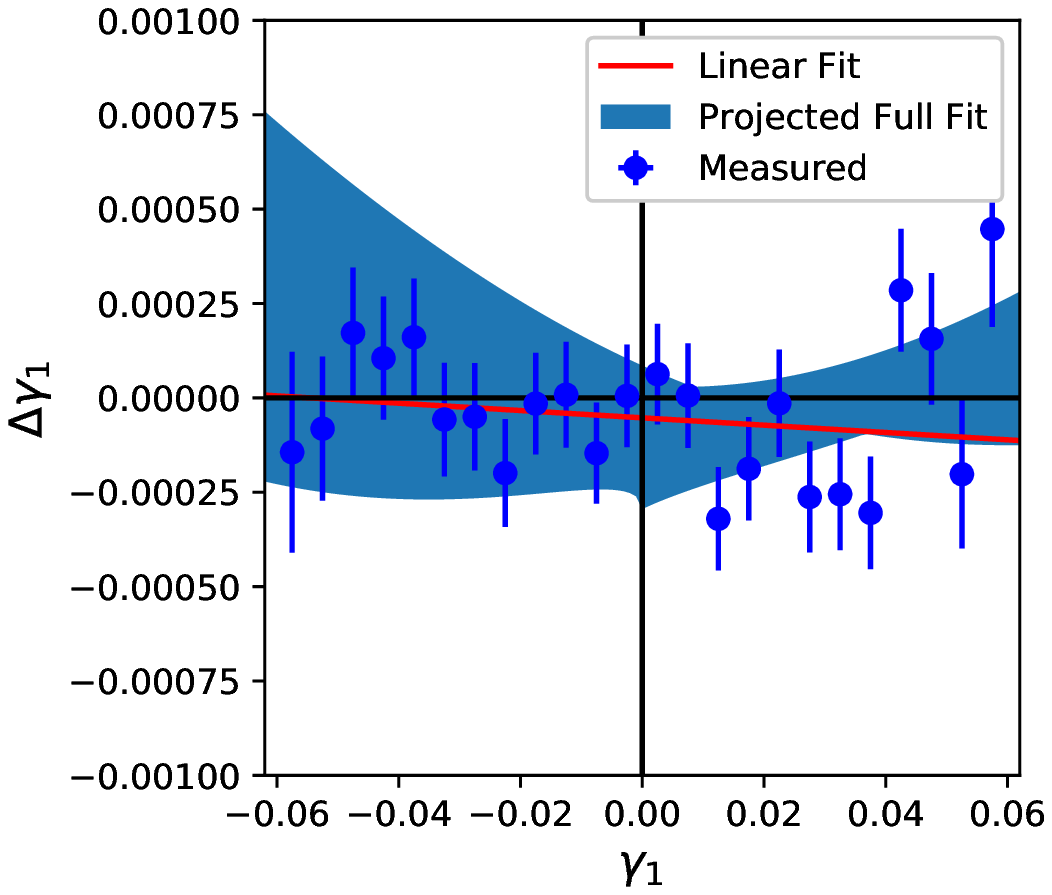}
\includegraphics[width=0.49\columnwidth]{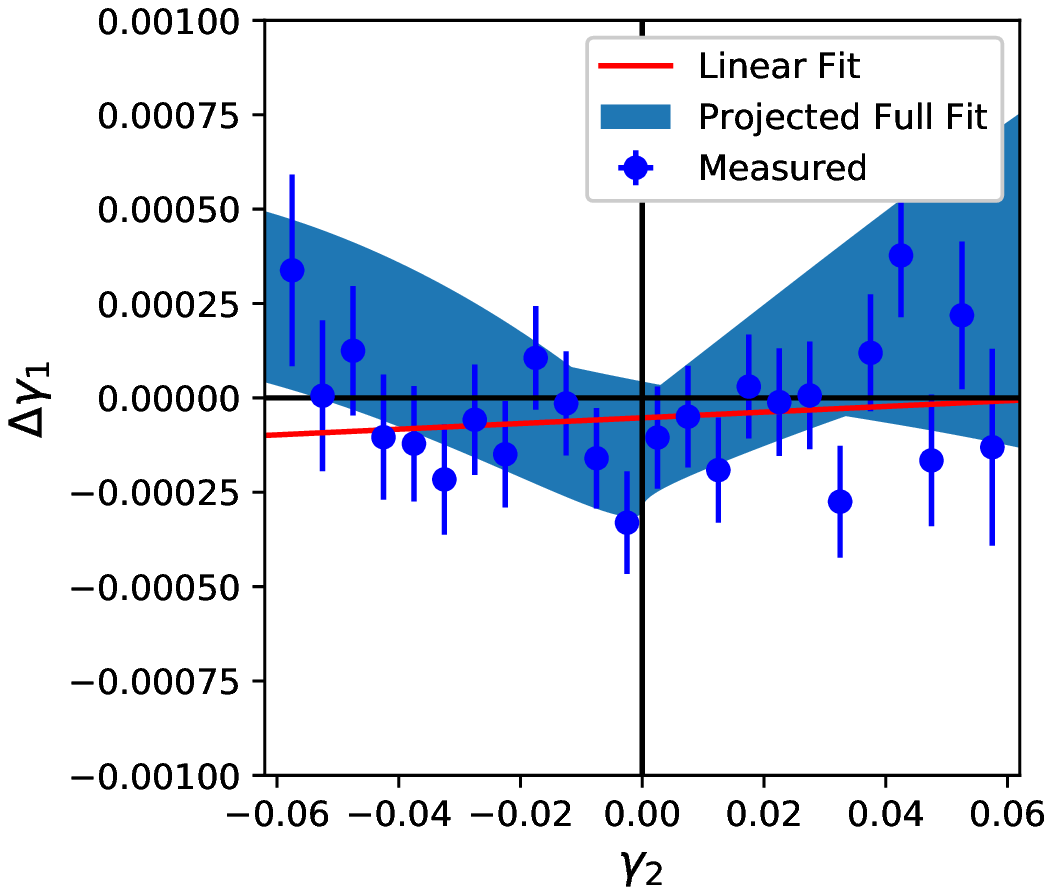}\\
\includegraphics[width=0.49\columnwidth]{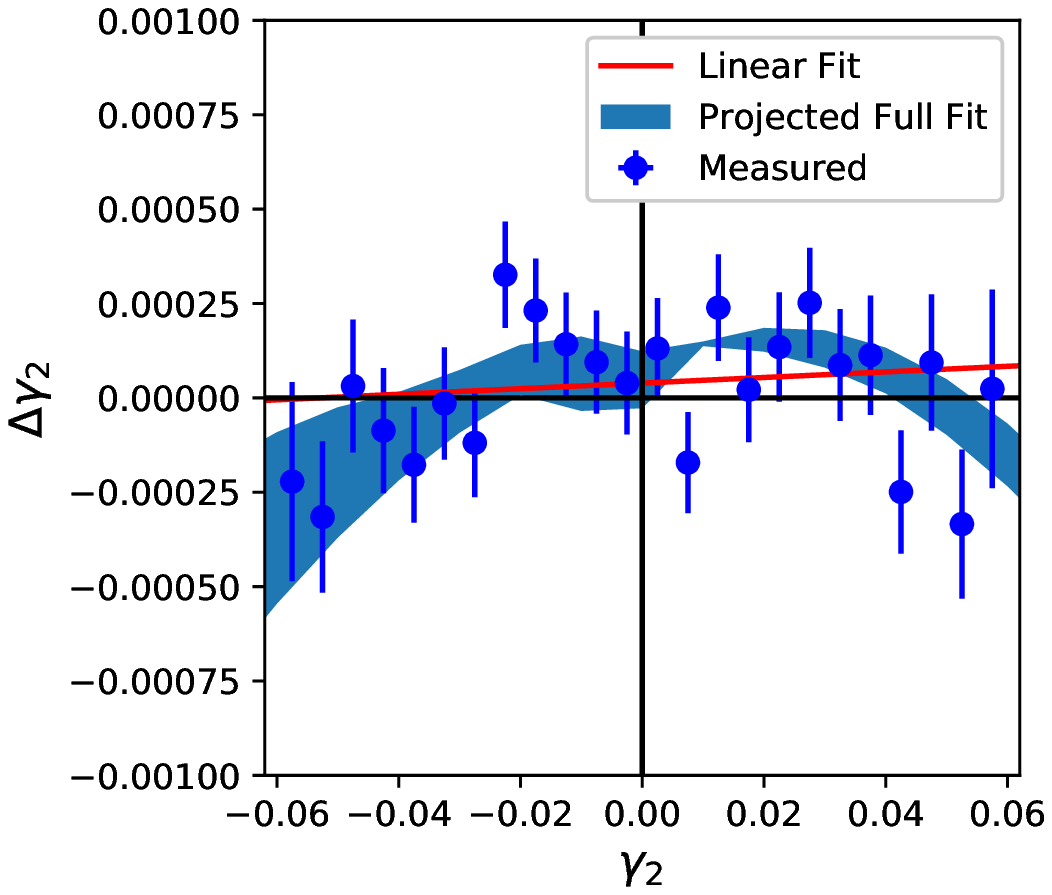}
\includegraphics[width=0.49\columnwidth]{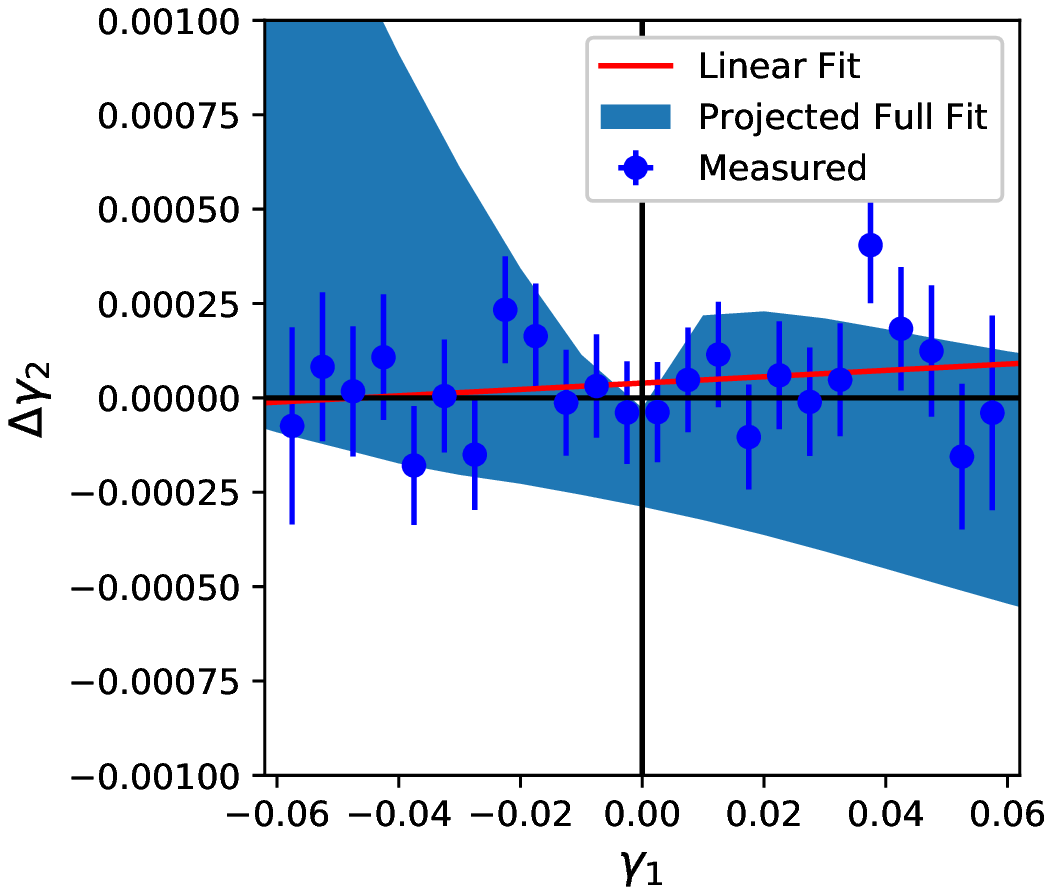}\\
\includegraphics[width=0.49\columnwidth]{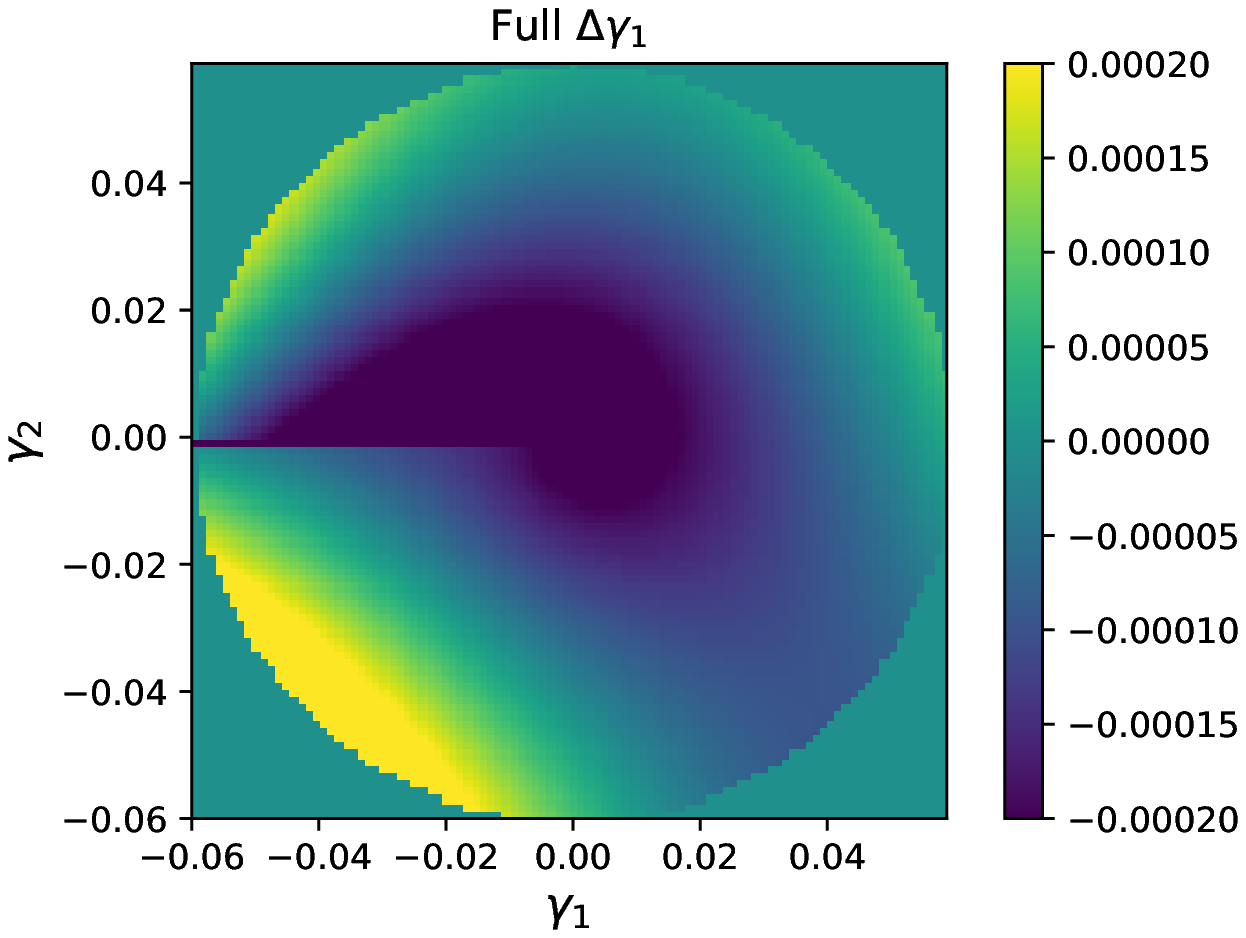}
\includegraphics[width=0.49\columnwidth]{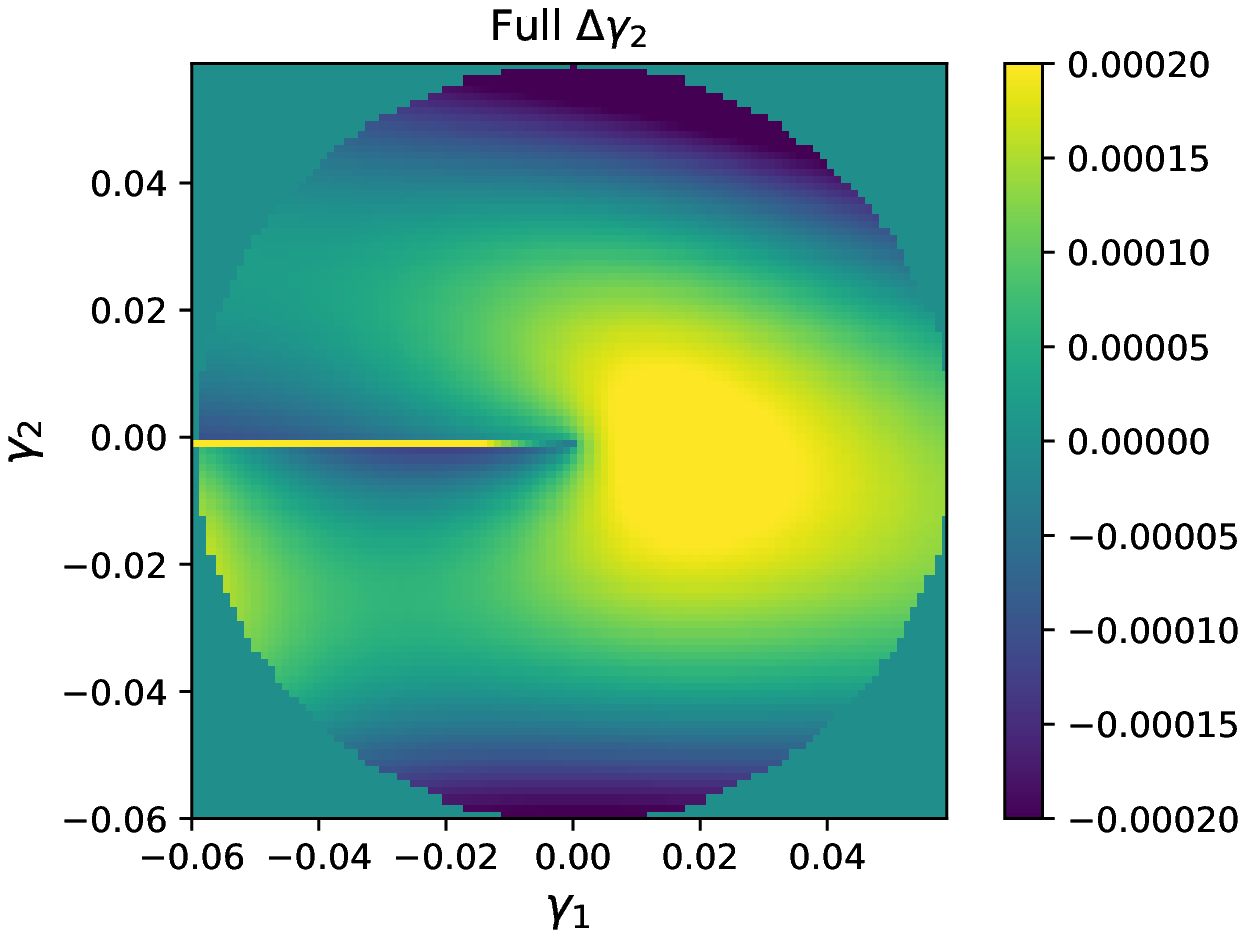}\\
 \caption{Regression of differences between measured and true shear values in the the $\Delta\gamma_1$ vs. $\gamma_1$; $\Delta\gamma_1$ vs. $\gamma_2$; $\Delta\gamma_2$ vs. $\gamma_2$; and $\Delta\gamma_2$ vs. $\gamma_1$; for the HSM \citep{HS,M} method using the `KSB' PSF correction method; using {\tt GalSim} \citep{galsim} {\tt demo2} using all terms up to second order (equation \ref{eeqfullfit}). In each plot we show the fit projected into each plane (blue regions lines), and in all cases a linear fit is shown (red lines). The bottom two plots show the $(\gamma_1,\gamma_2)$ plane with colour showing the amplitude of $\Delta\gamma_1$ (left) and $\Delta\gamma_2$ (right).} 
\label{fullfit}
\end{figure*}

In Figure \ref{fullfit} we show this full model fit to the calibration data for the HSB `KSB' method described in Section \ref{Test on a real method}. It can be seen that the surface fit to the $(\Delta\gamma_1,\Delta\gamma_2)$ plane is significantly more complex than that described by the fit only including integer terms, but that it is also a better fit to the data, capturing several features that were not well-fit by the integer terms' behaviour only. We find a reduced $\chi^2_R-1={\rm min}[\chi^2/(N_{\rm gal}-N_{\rm par})]-1=-4\times 10^{-5}$ where $N_{\rm par}=20$; which represents an improvement over the integer fit although both are good fits to the data. We find that the best-fit values of the full set of parameters are 
$m^R_0=0.012$, $m^R_4=0.006$, $m^R_2=0.009$,  $m^R_{-2}=0.089$, $m^R_6=0.043$, $c_1=-0.034$, $c_2=-0.019$, $m^R_1=-0.003$, $m^R_3=-0.002$, $m^R_{-1}=-0.044$, $m_5=-0.023$, $m^I_4=-0.026$, $m^I_2=0.011$, $m^I_{-2}=0.064$, $m^I_6=0.017$, $m^I_1=0.001$, $m^I_3=0.002$,$m^I_{-1}=-0.011$, $m^I_5=0.006$;
errors on each parameter are similar in order of magnitude to those in Table 1 (we exclude them here for clarity). 

It can be seen that by allowing flexibility in the model to include all terms up to second order that the best fit values can change significantly compared to the case of fitting only integer terms (see Table 1), in particular the additive bias values. This is a more general, higher-dimensional, feature that was observed in Section \ref{Test on Gaussian simulations} i.e. that  if a simplified model is fit to calibration data that the inferred values of the biases can be biased.

\subsection{The Impact of Quadratic-Order Half-Integer Terms on Cosmic Shear Power Spectra}
Taking all terms to linear order in bias the impact of half-integer terms on the cosmic shear power spectrum is to introduce the following 
\begin{eqnarray}
   \delta C_{\ell;\alpha\beta}&=&2[
   m^R_{-1} C^{\gamma(\gamma^{3/2})}_{\ell;\alpha\beta}+
   m^R_1 C^{\gamma(\gamma^{1/2})}_{\ell;\alpha\beta}+
   m^R_3 C^{\gamma(\gamma^{*,1/2})}_{\ell;\alpha\beta}+
   m^R_5 C^{\gamma(\gamma^{*,3/2})}_{\ell;\alpha\beta}\nonumber\\
   &+&m^R_0C^{\gamma(\gamma^{3/2}\gamma^{*,1/2})}_{\ell;\alpha\beta}+
   m^R_1C^{\gamma(\gamma\gamma^{*,1/2})}_{\ell;\alpha\beta}+
   m^R_2C^{\gamma(\gamma^{1/2}\gamma^{*,1/2})}_{\ell;\alpha\beta}+
   m^R_3C^{\gamma(\gamma^{1/2}\gamma^{*})}_{\ell;\alpha\beta}+
   m^R_4C^{\gamma(\gamma^{1/2}\gamma^{*,3/2})}_{\ell;\alpha\beta}]+{\mathcal O}(m^2)
\end{eqnarray}
where $C^{X(Y)}_{\ell;\alpha\beta}$ corresponds to the correlation $\langle XY^*\rangle$.

The impact of half-integer terms on the cosmic shear power spectrum is complicated and not simply a matter of taking a suitable exponent of the power spectrum or the bispectrum. In general, for a tomographic redshift bin $\alpha$, the spherical harmonic representation of the shear can be written:
\begin{align}
    \label{eq:convspehe}
    \widetilde{\gamma}_{\alpha; \ell m} &= \frac{1}{\ell(\ell+1)}\sqrt{\frac{(\ell+2)!}{(\ell-2)!}} 4\pi i^\ell \int_0^{\chi_{\rm lim}} {\rm d}\chi W_{\alpha}(\chi) \int_0^\infty \frac{{\rm d}^3 k}{(2\pi)^3}j_\ell(k\chi){}_2Y^*_{\ell m}(\boldsymbol{\hat{k}})\widetilde{\delta}(\boldsymbol{k}, \chi),
\end{align}
where $\widetilde{\delta}$ is the matter overdensity of the Universe, and the spatial momentum vector $\boldsymbol{k}$ has magnitude $k=|\boldsymbol{k}|$; $W_{\alpha}$ is given in equation (\ref{Wa}).

To construct the power spectrum corresponding to the cross-correlation between shear and a half-integer power $\langle\gamma\gamma^{*,1/2}\rangle$ for example would therefore require a calculation that involved the following correlation in $\widetilde{\gamma}_{i; \ell m}$
\begin{eqnarray}
    \delta C^{\gamma\gamma^{1/2}}_{\ell;\alpha\beta}\propto&& \frac{1}{2\ell+1}\sum_m 
    \left[\int_0^{\chi_{\rm lim}} {\rm d}\chi W_{\alpha}(\chi) \int_0^\infty \frac{{\rm d}^3 k}{(2\pi)^3}j_\ell(k\chi){}_2Y^*_{\ell m}(\boldsymbol{\hat{k}})\widetilde{\delta}(\boldsymbol{k}, \chi)\right]\nonumber\\
    &&\left[\int_0^{\chi_{\rm lim}} {\rm d}\chi W_{\beta}(\chi) \int_0^\infty \frac{{\rm d}^3 k}{(2\pi)^3}j_\ell(k\chi){}_2Y^*_{\ell m}(\boldsymbol{\hat{k}})\widetilde{\delta}(\boldsymbol{k}, \chi)\right]^{*,1/2},
\end{eqnarray}
a similar expression would exist to third order for the bispectrum. It can be seen that the standard factoring of the power spectrum into kernels multiplied by a matter overdensity power spectrum is not possible in this case. 

\begin{figure*}
\centering
\includegraphics[width=0.49\columnwidth]{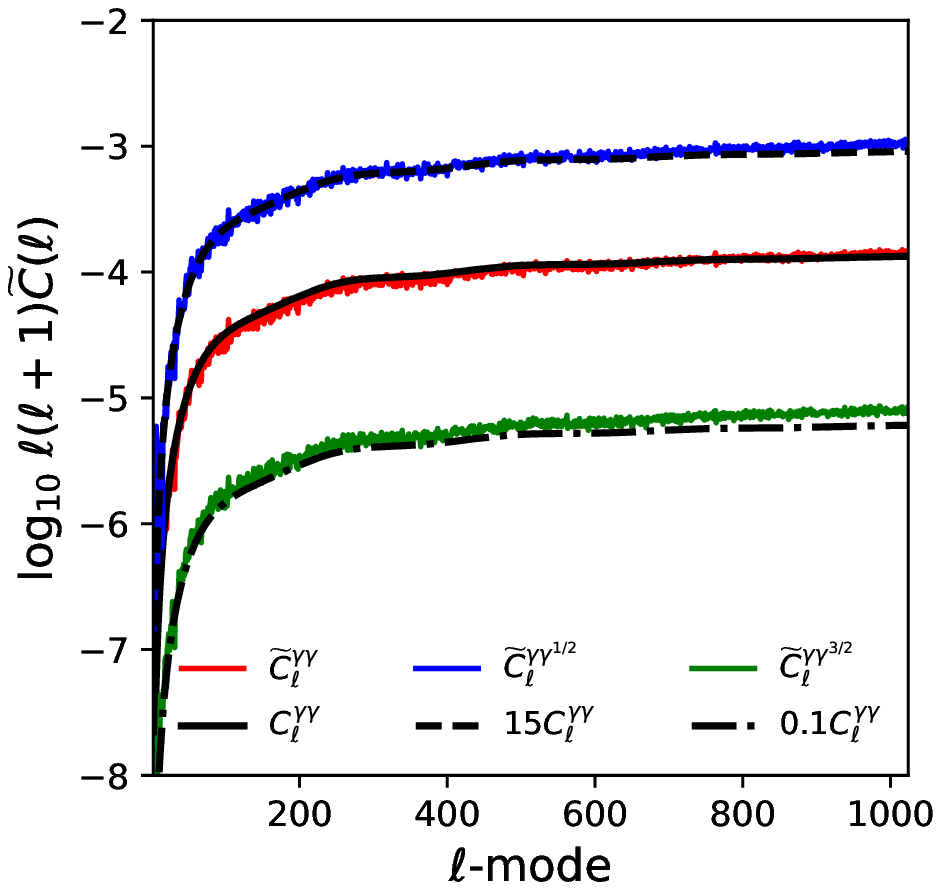}
\includegraphics[width=0.49\columnwidth]{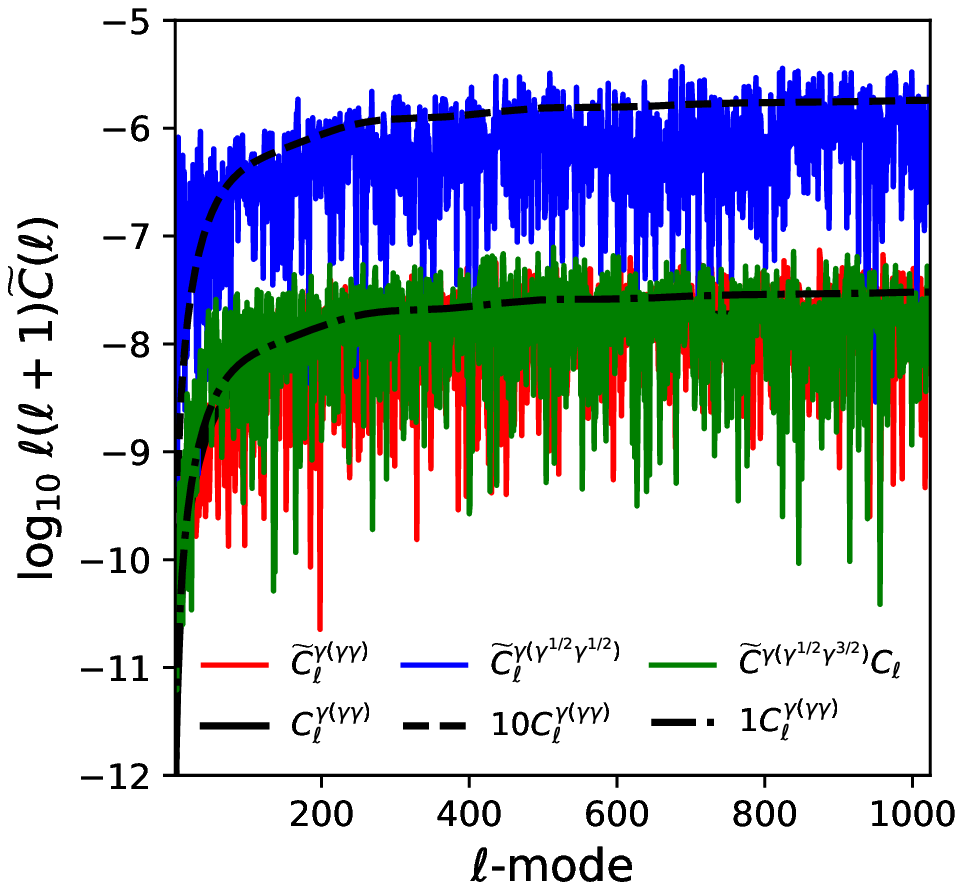}
 \caption{Left: the power spectrum of a Gaussian realisation of a true shear field with $C^{\gamma\gamma}_{\ell}$ (red line) compared to the input model (black solid line), and the cross-correlation power spectra $C^{\gamma\gamma^{1/2}}_{\ell}$ (blue) and $C^{\gamma\gamma^{3/2}}_{\ell}$ (green) compared to a simple amplitude scaling of $AC^{\gamma\gamma}_{\ell}$ with $A=6$ and $A=0.035$ (dashed and dotted lines respectively). Right: the impact on the power spectrum of the observed bispectrum of Gaussian realisation of a true shear field with $C^{\gamma\gamma}_{\ell}$. The red line shows $C^{\gamma(\gamma\gamma)}_{\ell}$, the blue line $C^{\gamma(\gamma\gamma^{1/2})}_{\ell}$ and the green line $C^{\gamma(\gamma\gamma^{3/2})}_{\ell}$. The solid, dashed and dot-dashed lines show $AC^{\gamma(\gamma\gamma)}_{\ell}$ with $A=1, 10$ and $1$ respectively.}
\label{power}
\end{figure*}

To understand the magnitude of such effects, rather than calculate such terms explicitly we instead create Gaussian random fields consistent with a true shear field with a power spectrum described in equation  (\ref{eq:powerspecdef}), for a single tomographic bin, with cosmological parameters set to \emph{Planck} maximum likelihood values \citep{planck} and a \cite{des1} number density \cite[see][for details]{ad2}. We then take half-integer powers of the generated shear field and compute $C^{\gamma\gamma^{1/2}}_{\ell}$ and $C^{\gamma\gamma^{3/2}}_{\ell}$. We show the result in Figure \ref{power}. We make an ansatz that the functional form of such terms will be a simple amplitude scaling of the shear power spectrum $AC^{\gamma\gamma}_{\ell}$. We find that this ansatz does accurately fit the power spectra with $A=15$ and $A=0.1$ for $C^{\gamma\gamma^{1/2}}_{\ell}$ and $C^{\gamma\gamma^{3/2}}_{\ell}$ respectively. We also find that $C^{\gamma\gamma^{1/2}}_{\ell}=C^{\gamma\gamma^{*,1/2}}_{\ell}$ and $C^{\gamma\gamma^{3/2}}_{\ell}=C^{\gamma\gamma^{*,3/2}}_{\ell}$ to a good approximation.

Similarly we find that the second order terms are approximately fit with amplitude scaling of the contribution to the power spectrum from the integer-shear bispectrum $AC^{\gamma(\gamma\gamma)}_{\ell}$ with $A=10$ and $1$ for $C^{\gamma(\gamma\gamma^{1/2})}_{\ell}$ and $C^{\gamma(\gamma\gamma^{3/2})}_{\ell}$ respectively. We emphasise that these tests are only using Gaussian random fields and future work should test if such anstaz apply for more realistic simulations.

\subsection{Discussion of Half-Integer Terms} 
\label{halfDiscussion} 
We find in general that the inclusion of half-integer terms is much more complex than integer terms for several reasons. First that the amplitude of such effects may be large -- because shear is less than unity, the square root is larger than the original quantity; secondly the impact on the change in $\gamma_1$ and $\gamma_2$ is highly non-trivial and not open to analytic derivation, as was the case with the integer terms; thirdly the power spectrum and bi-spectrum terms are also difficult to compute analytically because they do not factor in the way that integer terms do.

We find for the HSB method that the fit including half-integer terms is in general a better fit to the projected $\Delta\gamma_i$ vs. $\gamma_j$ planes, but that the shape is complex. Furthermore all $20$ free parameters corresponding to the real and imaginary parts of $m_s$ (with $s=-2$--$5$) plus the additive terms are non-zero. 

Regarding the power spectrum, we find that a simple amplitude scaling of the shear power spectrum is sufficient to capture the behaviour of linear order half-integer terms, and that a simple scaling of bi-spectrum is sufficient to capture quadratic order half-integer terms. If the ansatz for the bispectrum terms is correct then applying $k$-cut cosmic shear should, in a similar manner to the quadratic integer terms, remove sensitivity to such terms. 

However, the half-integer terms also introduce new half-integer order bias parameters that lead to an overall change in power spectrum of 
\begin{eqnarray}
    \delta C_{\alpha\beta} \simeq 2[15(m^R_1+m^R_3)C^{\gamma\gamma}_{\ell;\alpha\beta}+0.1(m^R_{-1}+m^R_5)C^{\gamma\gamma}_{\ell;\alpha\beta}],
\end{eqnarray}
assuming the amplitude scaling ansatz is correct.

\section{Discussion}
\label{Discussion}
In this section we address the overall impact of quadratic terms including both integer and half-integer terms. We have found that quadratic terms introduce additional bispectra to the cosmic shear power spectrum. For integer terms these are multiplied by $m^R_2$, $m^R_{-2}$ and $m^R_6$, in the half-integer case by $m^R_0$, $m^R_1$, $m^R_2$, $m^R_3$ and $m^R_4$. If no mitigation approach is taken cosmic shear power spectra will depend on bispectra that include contributions from all these fields. However, as discussed in Section \ref{Discussion} bispectrum terms can be removed by making scale-dependent cuts to the data vectors. Therefore the important contributions to the power spectrum, up to second-order in shear, are 
\begin{eqnarray}
    \delta C_{\ell;\alpha\beta} \simeq 2[m^R_0+15(m^R_1+m^R_3)+0.1(m^R_{-1}+m^R_5)]C^{\gamma\gamma}_{\ell;\alpha\beta},
\end{eqnarray}
this differs from the usual expression that only includes the $m^R_0$ term. For Stage-IV experiments requirements on $m^R_0$ have been set such that if $\sigma(m^R_0)\leq 0.01$ cosmological constraints should be unbiased \citep[see][]{K20,2022arXiv220301460C}. We find that this requirement should be on $\sigma[m^R_0+15(m^R_1+m^R_3)+0.1(m^R_{-1}+m^R_5)]$ i.e. that $m^R_0$, $m^R_1$, $m^R_3$, $m^R_{-1}$ and $m^R_5$ need to be determined with an overall summed uncertainty of $0.01$. Assuming equipartition between the five terms would suggest that each term would need to be constrained to $4\times 10^{-3}$. Of the five terms the large prefactor multiplying $(m^R_1+m^R_3)$ suggested that this is the most important to constraint. Only $m^R_0$ has been determined in experiments to date. We note that whilst the the metacalibration method \citep[see e.g.][]{2017arXiv170202600H,2022arXiv220308845Y} currently does not correct for non-scalar multiplicative bias fields that is could be generalised to do so. 

Finally, we note that the $20$ dimensional parameter fit is very large, and there is considerable flexibility in the model and degeneracy. Future work should investigate how to incorporate any  uncertainties into cosmological inference, and how to place appropriate priors on each multiplicative bias.

\section{Conclusions}
\label{Conclusions}
In this paper we revisit potential biases in cosmic shear power spectra caused by quadratic multiplicative bias terms. We find that such biases can be measured by performing general regression on $\Delta\gamma_i$ as a function of $\gamma_1$ and $\gamma_2$. 

We find that terms of integer power in shear and first order in bias impact on the power spectrum is the same as that caused by the reduced shear approximation, an additional bispectrum dependency, except multiplied by $2(m_2+m_{-2}-m_6)$. Ignoring quadratic terms can lead to biases in cosmological of up to $2(m_2+m_{-2}-m_6)0.4\sigma$. A reasonable requirement on the amplitude of these terms is $|m_2+m_{-2}-m_6|\leq 0.031$. If one ignores quadratic terms entirely and instead fits only a linear dependence to calibration data then this can in addition cause spurious multiplicative and additive biases. 

Investigating terms of half-integer power in shear we find new linear order terms, and new quadratic terms. The linear order terms introduce power spectrum changes proportional to the shear power spectrum, and the quadratic order terms introduce power spectrum changes proportional to the bispectrum. We find that a requirement of $\sigma[m^R_0+15(m^R_1+m^R_3)+0.1(m^R_{-1}+m^R_5)]\leq 0.01$ is needed for cosmological constraints from Stage-IV experiments to be unbiased.

In future therefore Stage-IV dark energy experiment should therefore seek to measure and minimise quadratic bias terms, and investigation should be done to apply Bayesian parameter estimation and model comparison to find the best model and parameter combination to apply to calibration data.

\vspace{-0.3cm}
\acknowledgements
\small{{\emph{Acknowledgements:} TDK acknowledges funding from the EU's Horizon 2020 programme, grant agreement No 776247. ACD acknowledges funding from the Royal Society. We thank Henk Hoekstra for comments on an early draft. We thank Andy Taylor, Alex Hall and Giuseppe Congedo for discussion and comments.}}

\bibliographystyle{mnras}
\bibliography{sample.bib}

\label{lastpage}
\end{document}